\documentstyle[prd,aps,floats,epsf]{revtex}

\newcommand{\eq}[1]{equation~(\ref{#1})}
\newcommand{\Eq}[1]{Equation~(\ref{#1})}
\newcommand{\eqs}[2]{equations~(\ref{#1}) and~(\ref{#2})}

\newcommand{\fig}[1]{fig.~(\ref{#1})}
\newcommand{\Fig}[1]{Fig.~(\ref{#1})}
\newcommand{\figs}[2]{figs~(\ref{#1}) and~(\ref{#2})}

\newcommand{\mpl}{m_{\mbox{\tiny pl}}}

\newcommand{\dphigi}{\delta\phi^{\mbox{\tiny(gi)}}}

\newcommand{\be}{\begin{equation}}
\newcommand{\ee}{\end{equation}}

\newcommand{\bea}{\begin{eqnarray}}
\newcommand{\eea}{\end{eqnarray}}

\def\alphaT {\alpha_{,\eta}}
\def\betaT {\beta_{,\eta}}
\def\PT {\phi_{,\eta}}
\def\ST {\psi_{,\eta}}

\def\alphaTT {\alpha_{,\eta\eta}}
\def\betaTT {\beta_{,\eta\eta}}
\def\PTT {\phi_{,\eta\eta}}
\def\STT {\psi_{,\eta\eta}}

\def\alphaZ {\alpha_{,\zeta}}
\def\betaZ {\beta_{,\zeta}}
\def\PZ {\phi_{,\zeta}}
\def\SZ {\psi_{,\zeta}}

\def\alphaZZ {\alpha_{,\zeta\zeta}}
\def\betaZZ {\beta_{,\zeta\zeta}}
\def\PZZ {\phi_{,\zeta\zeta}}
\def\SZZ {\psi_{,\zeta\zeta}}

\def\betaTZ {\beta_{,\eta\zeta}}
\def\ha {\frac{1}{2}}

\begin{document}
\draft

%
%

%
\renewcommand{\topfraction}{0.99}
\renewcommand{\bottomfraction}{0.99}
\twocolumn[\hsize\textwidth\columnwidth\hsize\csname 
@twocolumnfalse\endcsname

\title{Gravity, Parametric Resonance and Chaotic Inflation}
\author{Richard Easther$^{1}$ and Matthew Parry$^{2}$}
\address{   $^{1}$ Department of Physics,  Brown University, 
Providence, RI  02912, USA. \\
$^{2}$  Theoretical Physics Group, Blackett Laboratory, Imperial
College, Prince Consort Rd, London, SW7 2BZ, UK.}
\date{October 19, 1999}
\maketitle
\begin{abstract}
           
  We investigate the possibility that nonlinear gravitational effects
  influence the preheating era after inflation.  Our work is based on
  numerical solutions of the inhomogeneous Einstein field equations,
  and is free of perturbative approximations.  The one restriction we
  impose is to limit the inhomogeneity to a single spatial direction.
  We compare our results to perturbative calculations and to solutions
  of the nonlinear field equations in a rigid (unperturbed) spacetime,
  in order to isolate gravitational phenomena.  We consider two types
  of initial conditions: where only one mode of the field perturbation
  has a non-zero initial amplitude, and where all the modes begin with
  a non-zero amplitude.  Here we focus on preheating following
  inflation driven by a scalar field with a quartic potential.  We
  confirm the broad picture of preheating obtained from the nonlinear
  field equations in a rigid background, but gravitational effects
  have a measurable impact on the dynamics for both sets of initial
  data. The rigid spacetime results predict that the amplitude of a
  single initially excited mode drops rapidly after resonance ends,
  whereas in the relativistic case the amplitude is roughly constant.
  With all modes initially excited, the longest modes in the
  simulation grow much more rapidly in the relativistic calculation
  than with a rigid background.  However, we see no evidence for the
  sort of gravitational collapse associated with the formation of
  primordial black holes.  The numerical codes described here are
  easily extended to more complicated resonant models, which we will
  examine in the future.

\end{abstract}

]


\section{Introduction} 

For nearly a decade it has been known that coherent processes at the
end of inflation can drive explosive particle production
\cite{TraschenET1990a} via parametric resonance, leading to an era of
{\em preheating\/} in the post-inflationary universe.  Earlier
assumptions about the transfer of energy from the inflaton to
``regular'' matter have been overturned, and the phenomenology of the
post-inflationary universe is thus much richer than previously
suspected. However, the corollary of such interesting behavior is
understanding preheating has been a long and
arduous task%
\cite{PolarskiET1992a,%
KofmanET1994a,%
ShtanovET1995a,%
ProkopecET1996a,%
Kaiser1996a,%
KhlebnikovET1996a,%
KhlebnikovET1997a,%
KhlebnikovET1997b,%
KofmanET1997a,%
GreeneET1997a,%
GreeneET1997b,%
KodamaET1996a,%
HamazakiET1996a,%
KolbET1996a,%
NambuET1996a,%
Kaiser1997a,%
Kaiser1997b,%
NambuET1998a,%
TaruyaET1998a,%
KolbET1998a,%
BassettET1998a,%
FinelliET1998a,%
ParryET1998a,%
BassettET1999a,%
JedamzikET1999a,%
BassettET1999b,%
BassettET1999c}.
In particular, the problem is intrinsically nonlinear and takes place
far from thermal equilibrium.  Moreover, employing simplifying
assumptions that are often useful in other circumstances can suppress
crucial features of the preheating era, and when they are discarded
the resulting predictions can change dramatically.

The simplest model to exhibit parametric resonance is chaotic
inflation driven by the quartic potential,
\be
V(\phi) = \frac{\lambda}{4} \phi^4.    \label{phi4pot}
\ee
The $\phi$ particles themselves are produced parametrically, so there
is no need to couple the inflaton to other bosons, which is a
prerequisite for resonance when $V(\phi) = m^2\phi^2/2$.  By shifting
to conformal time and rescaling the field, all explicit dependence on
the expanding background can be removed from the field's equation of
motion.  Consequently, the analysis of parametric resonance for
$\lambda \phi^4$ is effectively performed in Minkowski space, and is
thoroughly discussed by Greene {\em et al.}  \cite{GreeneET1997a}.
The instability diagram contains a single, narrow resonance band and,
to first order, co-moving modes which are initially in resonance
appear to remain there for ever. Thus stochastic resonance, associated
with a given wavelength passing through multiple resonance bands, does
not occur.  However, when the back reaction of the created particles on
the field evolution is incorporated the resonant growth terminates
\cite{KhlebnikovET1996a,GreeneET1997a}.

The inflaton field, $\phi$, is described by a homogeneous background
part, $\phi_0(t)$, and an inhomogeneous part, $\delta\phi(t,x^i)$,
which is usually expressed in terms of its Fourier modes,
$\delta\phi_k$. Modes with wavenumbers, $k$, inside the resonance band
grow exponentially, while other modes have oscillatory
solutions. Because of the amplification of some of the $\delta\phi_k$,
the ``matter'' develops a substantial inhomogeneous component. The
background spacetime is usually assumed to be rigid, and therefore
unperturbed.

A full treatment of parametric resonance must include the
inhomogeneous metric induced by the inhomogeneous matter.
Incorporating metric perturbations into the analysis of parametric
resonance has been the subject of considerable recent attention
\cite{KodamaET1996a,%
HamazakiET1996a,%
NambuET1996a,%
BassettET1998a,%
FinelliET1998a,%
ParryET1998a,%
BassettET1999a,%
JedamzikET1999a,%
BassettET1999b,%
BassettET1999c}.
Perturbation theory cannot give a full description of resonance, since
once the $\delta\phi_k$ grow large, mode-mode couplings become
significant. Consequently, a better understanding is obtained from a
direct (numerical) solution of the nonlinear field equations. These
equations are nonlinear due to the higher order terms in the
potential, which are a necessary condition for resonance.  However, in
these simulations the background spacetime is an unperturbed FRW
universe, so gravitational effects are implicitly ignored.  To capture
both the full nonlinear physics of preheating and any gravitational
effects that influence the dynamics, we solve the evolution equations
for the fields and the spacetime background.  Our only {\it a
  priori\/} restriction is to confine the inhomogeneity to a single
spatial dimension, which reduces the differential equations we must
solve to a $1+1$ dimensional system.

Finally, it has been argued that second order terms in cosmological
perturbation theory contribute to the overall dynamics via back
reaction effects \cite{MukhanovET1997b,AbramoET1997a}.  In the context
of preheating the one calculation which explicitly includes second
order effects showed that they did not significantly affect the
dynamics \cite{JedamzikET1999a}.\footnote{However, a very recent paper
  by Bassett {\em et al.\/} takes issue with these conclusions.}
The broader issue of the back reaction of perturbations on
cosmological spacetimes can be addressed with our code, and we intend
to pursue this topic in the future.

In general, we can anticipate two possible outcomes. If gravitational
effects do not alter the dynamics the initial phase of the evolution
of the metric perturbations will reproduce the results of gauge
invariant perturbation theory%
\cite{KodamaET1996a,%
NambuET1996a,%
MukhanovET1992b,%
Mukhanov1988a,%
Hwang1996b}, 
and at later times, when the perturbative limit is no longer valid, we
will recover the results of the lattice simulations of the nonlinear
field equations in a rigid background.  On the other hand, if
gravitational effects do play an important role, we will see phenomena
that are not predicted by other approaches. 

Two possible gravitational effects have been identified by previous
work.  On small scales, the increased inhomogeneity induced by
resonance may lead to local gravitational collapse, and the formation
of primordial black holes \cite{BassettET1998a}.  Since they decay by
emitting Hawking radiation, any such holes would provide an additional
channel for thermalizing the post inflationary universe
\cite{GarciaBellidoET1996b}.  Many inflationary models do lead to the
formation of primordial black holes, provided the perturbation
spectrum they produce has more power at short scales than at COBE
scales \cite{CarrET1994a,GreenET1997a}.  However, if preheating
creates a large amount of inhomogeneity at small scales, it could lead
to primordial black hole formation in models where they would not
otherwise be expected. On scales much larger than the
post-inflationary Hubble length, it has been proposed that nonlinear
effects enhance metric perturbations \cite{BassettET1999a},
potentially undermining the usual inflationary predictions for
structure formation and the anisotropy of the microwave background.
Consequently, during our analysis we are particularly interested in
looking for evidence for both gravitational collapse at short length
scales, and the enhancement of metric perturbations at long length
scales. Finally, any changes in the initial resonance structure due to
the inclusion of gravitational perturbations would be revealed by
our simulations.

This paper extends the approach described by us in \cite{ParryET1998a}
(see also \cite{EastherET1999b,ParryET1999a}) in two ways. Firstly, we
generalize the allowed range of initial data so that $\delta\phi$ can
be an arbitrary function of position on the initial slice. Because of
this, we can analyze the preheating era after chaotic inflation driven
by a quartic potential, \eq{phi4pot}, with a realistic set of initial
conditions.  The further extension to models with more than one scalar
field is straightforward and poses no new technical challenges.
However, as the resonance provided by $\lambda \phi^4$ potential is
confined to a narrow band, it provides an excellent laboratory for
investigating both the perturbative and non-perturbative regimes,
before we turn our attention to more complex two-field models, or
single field models with more complex potentials.

In the following section, we assemble familiar results that describe
the evolution of a homogeneous universe after the end of inflation. We
then introduce the evolution equations for small perturbations, and
the nonlinear equations of motion for the inhomogeneous fields in an
unperturbed spacetime.  In Section 3, we derive the specific form
Einstein field equations for the inhomogeneous case, describe how we
fix the initial conditions; we then outline the numerical
algorithm we use to evolve the equations.  Section 4 describes the
results of our simulations.  We look at two different sets of initial
data. In the first, we start with a field perturbation that consists
of single Fourier mode, chosen to lie inside the resonance band.  This
choice facilitates comparison with perturbation theory, by minimizing
the effect of mode-mode couplings. The second, more realistic, choice
of initial data is to begin with a field perturbation where all modes
have a non-zero amplitude. Finally, in Section 5 we discuss our
results and the questions they raise.  For both the simulations, we
find noticeable differences between the approximate calculations and
the results obtained from solving the Einstein field equations.
However, these do not appear to be large enough to overturn the
overall picture obtained from lattice simulations of the nonlinear
field equations in a static background.

\section{Evolution Equations}
\subsection{Homogeneous Evolution }

For inflation in a flat, homogeneous universe, we have the general
result
\begin{eqnarray}
H^2  = \left(\frac{\dot{a}}{a}\right)^2 &=& \frac{\kappa^2}{3} \left[
   \frac{\dot{\phi}^2}{2}+ V(\phi)\right],\label{Hsqrd}
\\
\dot{H} &=& -\frac{\kappa^2}{2} \dot{\phi}^2 ,\label{Hdot}
\\
\ddot{\phi} &=& -3H\dot{\phi} -\frac{d V}{d \phi}. \label{phiddot}
\end{eqnarray}
where, as usual, $a(t)$ is the scale factor, and overdots stand for
differentiation with respect to the time, $t$. The gravitational
coupling is $\kappa^2 = 8\pi G = 8\pi/ \mpl^2$. For the special case
of a quartic potential, we make the transformation
\be
\eta = \int{\frac{1}{a(t)} dt},\qquad \chi = a \phi, 
\ee
after which the constraint, \eq{Hsqrd}, becomes
\be
a'^2 = \frac{\kappa^2}{3} \left[ \frac{1}{2} \left(\chi' - \chi
\frac{a'}{a} \right)^2 + \frac{\lambda \chi^4}{4}\right] ,
\label{Hconf}
\ee
and $\chi$ obeys
\be
\chi'' + \lambda \chi^3 - \frac{a''}{a} \chi =0.
\label{phiconf}
\ee
Primes refers to differentiation with respect to the conformal time,
$\eta$.  It is easy to show that the $\chi a'/a$ and $\chi a''/a
$ terms in \eqs{Hconf}{phiconf} respectively quickly become negligible
after the end of inflation \cite{GreeneET1997a}.

\subsection{Linear Perturbation Theory}

The onset of resonance can be studied perturbatively. In many cases,
the perturbative analysis of resonance is based on an expansion about
the homogeneous solution for the field, and assumes that the spacetime
background is unperturbed. However, including metric perturbations
(which couple to perturbations in the matter) does not make the
problem a great deal more difficult, and this is the approach we take
here.  Perturbation theory is most conveniently formulated using
Mukhanov's variable, $Q$
\cite{KodamaET1996a,NambuET1996a,MukhanovET1992b,Mukhanov1988a}, which
is related to (the $k$-th Fourier mode of) the gauge invariant
perturbation $\Phi_k$ \cite{MukhanovET1992b} by
\be
Q_k =\dphigi_k + \frac{\dot{\phi}}{H} \Phi_k, 
\label{Qvar}
\ee
where $\dphigi_k$ is the $k$-th Fourier mode of the gauge
invariant perturbation of $\phi$, and $\Phi$ is the gauge invariant
metric perturbation \cite{MukhanovET1992b}. With $n$ fields we have
$n$ distinct $Q$, which obey the coupled, second order equations
\cite{Hwang1996b}
\begin{eqnarray}
\ddot{Q}_i &+& 3H\dot{Q}_i + \frac{k^2}{a^2} Q_i + 
\sum_{j=1}^{n} \left\{ \frac{d^2 V}{\partial\phi_i \partial\phi_j}
\right. \nonumber \\
 && \left.
 - \kappa^2 \left[(3-\frac{\dot{H}}{H^2}) \dot{\phi}_i \dot{\phi}_j
+ \frac{1}{H}\frac{d}{dt}(\dot{\phi}_i \dot{\phi}_j )\right] \right\}
Q_j =0, \label{Qpert}
\end{eqnarray}
where the subscript on $Q$ now labels the field, and not the Fourier
mode.

The only relevant resonance band is
\cite{GreeneET1997a}
\be
k^2 = \left. c^2 \lambda a^2 \phi^2 \right|_{\dot{\phi}=0} 
\label{kband}
\ee
where
\be
\frac{3}{2}<c^2 <\sqrt{3}.
\ee
The time dependent quantities in \eq{kband} are all evaluated at an
(arbitrary) moment when the kinetic energy of the inflaton
vanishes. Strictly speaking, Greene {\em et al.\/}
\cite{GreeneET1997a} do not analyze \eq{Qpert}, since they assume that 
only the field is perturbed, and that there are no metric
perturbations.  However, the resonance structure is not significantly
changed by the presence of metric perturbations.

Because of the scaling properties of the solution, the values of $k$
which fall inside the resonance band are not time dependent. This
greatly simplifies the structure of resonance in this model, since
only a small number of modes experience any resonant amplification.
While we are interested the general question of nonlinear
gravitational effects during preheating, this property of $\lambda
\phi^4$ theory makes it an excellent testbed for exploring nonlinear
gravitational effects during resonance, and the tools we are
developing here will later be applied to more complicated models.

\subsection{Nonlinear Field Equations}

The next level of sophistication is to include all terms which are
nonlinear in the fields, but to assume that the background spacetime
is unperturbed.  In the context of $\lambda \phi^4$ theory, this
problem was first considered by Khlebnikov and Tkachev
\cite{KhlebnikovET1996a}.  The perturbative analysis of preheating
involves ordinary differential equations, but the nonlinear field
equations are partial differential equations which are typically
solved numerically on a spatial grid.  Resonance causes individual
modes to become large, so mode-mode couplings are significant and
first order perturbation theory breaks down.  The only nonlinear
couplings ignored by these solutions are those attributable to
gravity, and they thus provide a much more complete description of
resonance and preheating than perturbation theory. The purpose of this
paper is to find out whether this description is complete, and we do
this by comparing solutions of the nonlinear field equations with the
full relativistic calculation we describe in the next section.

We derive the nonlinear field equations by adding spatial gradient
terms to \eq{phiconf},
\be
\chi'' - \nabla^2 \chi + \lambda \chi^3 - \frac{a''}{a} \chi =0,
\label{chieofm}
\ee
where $\nabla = \partial_i$ and a prime denotes differentiation with
respect to conformal time. Working with the conformal field, $\chi$,
simplifies the equations. This equation includes gradients of $\chi$,
but the metric is implicitly assumed to be a FRW spacetime.  However,
if the spatial gradients are non-zero, the metric cannot be perfectly
homogeneous, so gravitational effects on preheating are ignored when
this equation is solved.

Ignoring perturbations, the post-inflationary expansion mimics a radiation
dominated universe, since on average the pressure and density of the
oscillating field obey $p=\rho/3$.  In this limit, $a(\eta) \propto \eta$,
and the $a''/a$ term rapidly becomes negligible after the oscillatory
phase begins. 

Our numerical code solves both the Einstein field equations and
\eq{chieofm}.  All the nonlinear terms in \eq{chieofm} also appear
in the relativistic equations. By comparing $\chi$ with the field
evolution obtained from the relativistic calculation we can isolate
gravitational effects during the preheating era.

To recover $\phi$ from our solution for \eq{chieofm} we divide $\chi$ by
$a$.  In order to obtain $a$, one could solve the ordinary differential
equations which give $a$, or assume an exact radiation dominated
expansion.  In practice we extracted $a$ by averaging the metric functions
derived from our numerical solutions of the full Einstein equations. Since
the average of the perturbation vanishes, we recover the background
solution in a self-consistent manner. However, we followed
\cite{KhlebnikovET1996a} and dropped the $a''/a$ term from \eq{chieofm}.

\section{Einstein Field Equations}

We now turn our attention to the explicit form of the Einstein field
equations which we will solve numerically. To reduce the computational
complexity of the problem, we assume that the universe has a planar
symmetry; the metric functions depend only on $t$ and $z$, and are
independent of $x$ and $y$. In \cite{ParryET1998a}, we examined
possible nonlinear gravitational effects after $m^2\phi^2$ inflation
using the metric
\be \label{metric}
ds^2 = dt^2 - A^2(t,z)\,dz^2 - B^2(t,z)\,(dx^2 + dy^2), 
\ee
which describes an inhomogeneous universe in which the $dx^2 + dy^2$
sections have zero spatial curvature.  In principle, we could retain
this metric here, but in practice it is simpler to work with a metric
for which the field equations reduce to the conformal time version of
the homogeneous system, rather than physical time which is the
homogeneous limit of \eq{metric}.  The oscillation time for the
homogeneous field is constant when expressed in conformal time, and
this property persists in the nonlinear system.  Thus the time-scale
governing the evolution does not change significantly during the
simulation, and we can use a constant timestep, simplifying our
numerical problem.

Specifically, a co-ordinate transformation from $t$ and $z$ to $\eta$
and $\zeta$ allows us to re-write \eq{metric} as:
\be\label{metric2}
ds^2 = \alpha^2(\eta,\zeta)\,(d\eta^2 - d\zeta^2) - 
\beta^2(\eta,\zeta)\,(dx^2 + dy^2).
\ee
With this metric, the non-trivial  components of the Einstein tensor
are
\begin{eqnarray}
G^{\,\eta}_{\,\:\eta} &=& \frac{1}{\alpha^2} \left[
2\frac{\betaZZ}{\beta} - \frac{\betaT^2}{\beta^2} +
\frac{\betaZ^2}{\beta^2} - 2\frac{\alphaT\betaT}{\alpha\beta} - 
2\frac{\alphaZ\betaZ}{\alpha\beta} \right],
\nonumber \\
G^{\,\eta}_{\,\:\zeta} &=& \frac{1}{\alpha^2} \left[2\frac{\betaTZ}{\beta} - 
2\frac{\alphaT\betaZ}{\alpha\beta} -
2\frac{\alphaZ\betaT}{\alpha\beta} \right],
\nonumber \\
G^{\,\zeta}_{\,\:\zeta} &=&   
\frac{1}{\alpha^2} \left[-2\frac{\betaTT}{\beta} - 
\frac{\betaT^2}{\beta^2} + 
\frac{\betaZ^2}{B^2} +2\frac{\alphaT\betaT}{\alpha\beta} + 
2\frac{\alphaZ\betaZ}{\alpha\beta} \right], 
\nonumber \\
G^{\,x}_{\,\:x} &=& \frac{1}{\alpha^2}
\left[-\frac{\alphaTT}{\alpha} -
\frac{\betaTT}{\beta} + \frac{\alphaZZ}{\alpha} + \frac{\betaZZ}{\beta} 
+ \frac{\alphaT^2}{\alpha^2} -
\frac{\alphaZ^2}{\alpha^2} \right], \nonumber \\
 G^{\,y}_{\,\:y}&=& G^{\,x}_{\,\:x}.
\end{eqnarray} 

The field Lagrangian is,
\be\label{lagr2}
{\cal L} = \ha \frac{\PT^2 - \PZ^2}{\alpha^2}  - V(\phi),
\ee
and yields the following equation of motion for $\phi$
\be\label{eomphi2}
\PTT = -2\frac{\betaT}{\beta}\,\PT + \PZZ + 
2\frac{\betaZ}{\beta}\,\PZ - V_{,\phi} \alpha^2.
\ee
The components of the energy-stress-momentum tensor are therefore 
\begin{eqnarray}
T^{\,\eta}_{\:\,\:\eta} &=&  \frac{1}{2\alpha^2}(\PT^2 + \PZ^2 ) + V,
\nonumber \\
T^{\,\eta}_{\:\,\:\zeta} &=& \frac{1}{\alpha^2} \PT\,\PZ,  \nonumber \\
T^{\,\zeta}_{\:\,\:\zeta} &=& -\frac{1}{2\alpha^2}
(\PT^2 + \PZ^2 ) + V,
\nonumber \\
T^{\,x}_{\:\,\:x} = T^{\,y}_{\:\,\:y} &=& -\frac{1}{2\alpha^2}
(\PT^2 - \PZ^2) + V. \nonumber
\end{eqnarray}

The corresponding field equations are 
\begin{eqnarray}
\label{eomalpha}
\alphaTT &=& -\frac{\alphaT\betaT}{\beta} + \frac{\alpha\betaT^2}{2\beta^2} 
+ \frac{\alpha\betaZZ}{\beta} -
\frac{\alpha\betaZ^2}{2\beta^2} - \frac{\alphaZ\betaZ}{\beta} + \nonumber \\
&&  \alphaZZ + 
\frac{\alphaT^2}{\alpha}- \frac{\kappa^2 \alpha}{2} 
\left[\ha \PT^2 -\frac{3}{2} \PZ^2  - V \alpha^2 \right], \\
\label{eombeta}
\betaTT &=& -\frac{\betaT^2}{2\beta} + \frac{\betaZ^2}{2\beta} + 
\frac{\alphaT\betaT}{\alpha} +
\frac{\alphaZ\betaZ}{\alpha} -\nonumber \\
&& \frac{ \kappa^2\beta}{2}
\left[\ha \PT^2 + \ha \PZ^2 - V \alpha^2 \right].
\end{eqnarray}
Finally, we have two constraints which must impose on the initial
data, but which are then satisfied at all later times as a consequence
of the field equations,
\begin{eqnarray}
\label{constr12}
&& \frac{\betaTZ}{\beta} - \frac{\alphaT\betaZ}{\alpha\beta} - 
\frac{\alphaZ\betaT}{\alpha\beta} = 
-\frac{\kappa^2}{2} \PT\,\PZ.
\\
\label{constr22}
&& -\frac{\betaT^2}{2\beta^2} - 
\frac{\alphaT\betaT}{\alpha\beta} + \frac{\betaZZ}{\beta} +
\frac{\betaZ^2}{2\beta^2} - \frac{\alphaZ\betaZ}{\alpha\beta} 
\nonumber \\
&=&
 -{\kappa^2 \over 2} \left[\ha \PT^2 + \ha \PZ^2  + V \alpha^2 \right].
\end{eqnarray}

\subsection{Initial Conditions}

To solve the equations, we must specify the initial conditions. These
cannot all be chosen arbitrarily, since the initial data must satisfy
the constraint equations.  We consider two possible scenarios. In the
first we excite a single mode and track its evolution using the full
nonlinear equations. In the second, all modes are initially excited.
Using the first approach it is easy to compare our results to a
perturbative calculation. However, beginning with a configuration in
which all the inhomogeneous modes have a small, non-zero amplitude is
more realistic. 

\subsubsection{A Single Excited Mode}

If we only wish to excite a single mode, we can pick initial
conditions which ensure that the constraints are automatically
satisfied, and thus do not need to be solved explicitly before
beginning the actual integration.  Following the approach of
\cite{ParryET1998a}, we set $\alpha(0,\zeta) = \beta(0,\zeta) = 1$,
$\phi(0,\zeta)=\phi_0$. \Eq{constr22} is solved if
\begin{eqnarray}
\alphaT(0,\zeta) &=& \frac{\kappa^2}{2C}\left( \frac{\PT^2}{2} +
V(\phi_0) \right) - \frac{C}{2}, \\
\betaT(0,\zeta) &=& 
\sqrt{\frac{\kappa^2}{3} \left( \frac{\langle\PT^2\rangle}{2} +
V(\phi_0)\right)}= C. 
\end{eqnarray}
Our choice of $C$ ensures $\langle \alphaT(0,\zeta) \rangle =
\langle \betaT(0,\zeta) \rangle$, where $\langle \cdots \rangle$ is a 
spatial average. The actual inhomogeneity is injected through the
inflaton velocity, which we choose to be
\be \label{pert1}
\PT(0,\zeta) =  \Pi + \epsilon \sin{\left(\frac{2\pi k \zeta}{Z}\right)},
\ee
where $Z$ is the length of our ``box'', and $\Pi$ is the average
initial velocity. However, since we have assumed that $\phi$, $\alpha$
and $\beta$ are independent of $\zeta$ on our initial slice we can
only excite multiple modes if we assume that they are all initially in
phase with each other.

\subsubsection{General Initial Data}                    

In the second scenario, where all the inhomogeneous modes are
initially excited, we want to allow their initial phases to be random.
With just one mode the phase is irrelevant, but demanding that all
modes begin with correlated phases would lead to an unacceptable loss
of generality.  In particular, we are interested in modeling the
Gaussian fluctuations predicted by inflation, which are defined in
part by having random phases.  In this case there is no simple and
obvious approach to solving the constraints. A general procedure does
exist to fix the initial data in $3+1$ dimensional numerical
relativity
\cite{York1982a}, but it is not necessarily easy to implement.  Since
we are interested in a perturbation that is initially small, we will
adopt the approach used by Goldwith and Piran
\cite{GoldwirthET1991b}, who simulate the onset of inflation in a
$1+1$ dimensional universe.  Their approach is to add a small amount
of radiation, with a spatial dependence chosen to ensure that the
overall density perturbation vanishes on the initial surface. Having
done this the constraints become simple to solve, while the radiation
introduced by this procedure always makes a trivial contribution to
the overall density, and quickly decays away.

We modify Goldwirth and Piran's approach slightly, and use an
uncoupled scalar field, $\psi$, instead of radiation. The energy
density of a massless free field drops as $a^{-6}$ in a homogeneous
universe, in contrast to radiation which scales as $a^{-4}$.  This
ensures that the contribution from $\psi$ to the overall density drops
more quickly than that of $\phi$, which behaves like radiation.

Including a second field in our code is simple. The field itself obeys 
the equation of motion,
\be\label{eompsi}
\STT = -2\frac{\betaT}{\beta}\,\ST + \SZZ + 
2\frac{\betaZ}{\beta}\,\SZ \, ,
\ee
and we modify \eqs{eomalpha}{eombeta} by adding $\psi_{,\zeta}$ and
$\psi_{,\eta}$ terms analogous to the derivatives of $\phi$ that are
already there. Most importantly for us, the revised constraints are
\begin{eqnarray}
\label{constr12a}
&& \frac{\betaTZ}{\beta} - \frac{\alphaT\betaZ}{\alpha\beta} - 
\frac{\alphaZ\betaT}{\alpha\beta}  \nonumber \\
&=& -\frac{\kappa^2}{2}\left( \PT\,\PZ + \ST\,\SZ \right),
\\
\label{constr22a}
&& -\frac{\betaT^2}{2\beta^2} - 
\frac{\alphaT\betaT}{\alpha\beta} + \frac{\betaZZ}{\beta} +
\frac{\betaZ^2}{2\beta^2} - \frac{\alphaZ\betaZ}{\alpha\beta} 
\nonumber \\
&=&
 -{\kappa^2 \over 2} \left[\ha \PT^2 + \ha \PZ^2+ \ha \ST^2 + \ha \SZ^2+
  V \alpha^2 \right].
\end{eqnarray}
We solve these equations on the initial slice by choosing $\SZ$ and
$\ST$ so that the right hand sides of the constraints simplify, or
\begin{eqnarray}
\label{constr12b}       
 \PT\,\PZ + \ST\,\SZ  &=&0, \\
\label{constr22b}
 \ha \PT^2 + \ha \PZ^2+ \ha \ST^2 + \ha \SZ^2+
  V \alpha^2 &=& D^2.
\end{eqnarray}
Now   we are free to choose
\be
\alpha(0,\zeta) = \beta(0,\zeta) = 1 \, ,\qquad
\alphaT(0,\zeta) = \betaT(0,\zeta) = H  \label{initvals}
\ee
where $H = D \kappa /\sqrt{3}$.  Adding \eq{constr22a} to
\eq{constr22b} gives two coupled equations for $\ST$ and $\SZ$, which
we solve to find
\begin{eqnarray}
\ST &=& \frac{A \pm \sqrt{A^2 - 4B}}{2}, \\
\SZ &=& \frac{A \mp \sqrt{A^2 - 4B}}{2} ,
\end{eqnarray}
where $A^2 = 2 D^2 - 2V - (\PT + \PZ)^2$ and $B = -\PZ
\PT$, and we have used $\alpha(0,\zeta)=1$. We can choose either of
the two solutions.  Finally, for these to be real, we need $A^2 - 4
B>0$, which is equivalent to demanding
\be \label{Dmax}
D^2 \ge \max{\left[ \ha (|\PT| + |\PZ|)^2 + V\right]}.
\ee
We are free to choose $D^2$, subject to the above constraint. In
practice we want to minimize the impact of the extra field, so we set
it close to the lower bound.  However, our numerical scheme is
somewhat more stable if the inequality is not completely saturated,
and in practice we chose a value of $D$ somewhat larger than the
minimum allowable value. Finally, $\SZ$ and $\ST$ are periodic, as
required by the boundary conditions.

\subsubsection{Choice of Initial Data}
We need to specify $\phi$ and $\PT$ on the initial slice.  We
begin our simulations at the end of inflation, so $\phi$ is described
by small fluctuations overlaid on top of the homogeneous zero
mode. The fluctuations are quantum in origin, and are generated by the
usual inflationary mechanism.  In practice the spectrum will have some
scale dependence, especially towards the end of inflation when the
slow roll approximation breaks down. If we wished we could track the
perturbations from their inception during inflation, but we will
assume the spectrum is initially scale-free since our results do not
depend on the precise form of the spectrum.

Let $q = \langle \phi(0,\zeta) \rangle$ and $p = \langle \PT(0,\zeta) 
\rangle$, where $\langle \cdots \rangle$ denotes a spatial average; these
values are chosen to be those at the end of inflation.  We will write
the perturbation in $\phi$, for {\it any} time, as
\be\label{pert}
X = \beta \left ( \phi - \langle \phi \rangle \right ).
\ee
Since, the metric perturbations which are first order in $\eta$ vanish
when $\eta=0$, we can expand \eq{eomphi2} to obtain an equation for
the homogeneous mode and a linear equation for $X$, valid on the
initial slice when $\eta=0$.  Taking the Fourier transform of this
latter equation gives
\be\label{eompert}
X_{{\bf k},\eta\eta} + \left ( k^2 + H^2 + \langle V_{,\phi\phi} - \kappa^2 V
\rangle \right ) X_{{\bf k}} = 0 .
\ee
Thus for a short time near $\eta=0$ each $k$-mode of $X$ can be
approximated as a simple harmonic oscillator with a $k$-dependent
frequency,
\be
\omega_k^2 = k^2 + H^2 + \langle V_{,\phi\phi} - \kappa^2 V \rangle.
\label{omegak}
\ee
Our approach will be to treat $X(0,\zeta)$ as a free quantum field in
the vacuum state; this amounts to the initial choice
\begin{eqnarray}
X_{{\bf k}} &=& {1 \over \sqrt{2\omega_k}} \, |n| \, e^{2\pi i r},
\label{qic1} \\
X_{{\bf k},\eta} &=& -i \omega_k X_{{\bf k}}, \label{qic2}
\end{eqnarray}
where $n$ is a number taken randomly from a Gaussian distribution of
mean 0 and variance 1, and $r$ is taken randomly from the interval
$[0,1]$.  For \eqs{qic1}{qic2} to be applicable the adiabatic limit
must hold, namely $\omega_{k,\eta} \ll \omega_k^2$. A more rigorous
treatment is possible but our approach is adequate for fixing the
scale of the fluctuations.  Moreover, we emphasize that while this
choice of initial conditions is a good approximation when applied
prior to the onset of resonance, its underlying assumptions will be
strongly violated at later times.

We insert $\phi$ into \eq{omegak} to determine $\omega_k$, from which we
learn $X_{\bf k}$ and, ultimately, $\phi$. If we insist the
fluctuations are small\footnote{In a numerical scheme, requiring small
fluctuations leads to a constraint on the lattice spacing. This is
because the lattice provides a natural UV (and IR) cutoff to the
unrenormalized quantum field theory we have effectively introduced in
\eq{qic1} and \eq{qic2}.} 
we can solve this problem iteratively.  We approximate $\langle
V_{,\phi\phi} - \kappa^2 V \rangle$ by $V_{,\phi\phi}(q) - \kappa^2
V(q)$. The final complication is that $H$ also depends on $\phi$
through \eq{constr22b} and \eq{initvals}.  Our approach here is as
follows: $H^2 = D^2\kappa^2/3$, and $D^2=\rho_{\phi} + \rho_{\psi}$ is a
constant, so $D^2 = \langle \rho_{\phi} + \rho_{\psi} \rangle$.  We
approximate $\langle \rho_{\phi}\rangle$ by $p^2/2 + V(q)$ and
then {\it choose} an $\epsilon$ and insist $\langle \rho_{\psi}
\rangle = \epsilon \langle \rho_{\phi}\rangle$. Setting $\epsilon \ll
1$ ensures the initial energy density of $\psi$ is small and fixes $D$
and $H$. However, if $\epsilon$ is too small it will not satisfy 
\eq{Dmax}. This can only be checked when $\phi$ and $\PT$ have been
determined: if $\epsilon$ is too small and one must repeat the process
with a larger value.  In practice $\epsilon$ can be on the order of
${1 \over 10}$, and the contribution of $\psi$ rapidly becomes 
negligible as the universe expands.

Once $X(0,\zeta)$ and $X_{,\eta}(0,\zeta)$ have been found though, it
is a simple matter to obtain the initial conditions for $\phi$ and
$\PT$ using \eq{pert}:
\begin{eqnarray}
\phi(0,\zeta) &=& q + X(0,\zeta), \label{phic1} \\
\PT(0,\zeta) &=& p + X_{,\eta}(0,\zeta) - H X(0,\zeta). \label{phic2}
\end{eqnarray}

\subsection{Coordinate Transformation}

To transform between the ``conformal'' and ``physical'' co-ordinates,
we need $t=t(\eta,\zeta)$ and $z=z(\eta,\zeta)$. The transformation
that connects \eq{metric2} to \eq{metric} reduces to the following
pair of differential equations
\begin{eqnarray}
t_{,\eta} &=& \sqrt{t_{,\zeta}^2 + \alpha^2(\eta,\zeta)} , 
\label{tft}\\
z_{,\eta} &=& \frac{t_{,\zeta} z_{,\zeta}}{t_{,\eta}},\label{tfz}
\end{eqnarray}
from which we find
\begin{eqnarray}\label{Atranf}
A(t,z) = A(t(\eta,\zeta),z(\eta,\zeta)) &=& \frac{t_{,\eta}}{z_{,\zeta}} ,\\
\label{Btranf}
B(t,z) = B(t(\eta,\zeta),z(\eta,\zeta)) &=& \beta(\eta,\zeta),
\end{eqnarray}
and the specification is completed with the following set of initial
conditions, 
\begin{equation}\label{ics}
t(0,\zeta) = 0, \qquad z(0,\zeta) = \zeta.
\end{equation}
Since both $t$ and $z$ are transformed, slices of constant $\eta$ in
the conformal frame are not mapped directly to slices of constant $t$
in the physical frame. 

As we have already noted, the $(\eta,\zeta)$ co-ordinates are
convenient because the typical timescale for the field oscillations is
roughly constant in this frame. Moreover, the degree of spatial
variation is $\alpha(\eta,\zeta)$ is significantly less than in
the corresponding $A(t,z)$, which is enhances our numerical stability.  In
practice, we solve \eqs{Atranf}{Btranf} alongside with the field
equations themselves, thereby obtaining solutions for both forms of
the metric.

\subsection{Numerical Code}

The actual numerical code is essentially the same as that used in
\cite{ParryET1998a}. The simulations are evolved forwards in time on
an $N$ point spatial grid, with periodic boundary conditions. We use a
fourth-order differencing scheme to express \eqs{eomalpha}{eombeta} as
$N$ ordinary differential equations, and solve these using a
fourth-order Runge-Kutta integrator. Numerical noise is controlled
with a de-aliasing algorithm, which requires filtering Fourier
components with wavelengths less than 8 grid spacings at each
timestep, so our actual spatial resolution is effectively 16 times
larger than the separation of the individual points, $\delta \zeta$
\cite{Choptuik1991a,GottliebBK1}. We track the numerical accuracy of
the solutions by studying how closely the constraint equations are
satisfied, and have also compared solutions obtained directly from the
metric \eq{metric}, and those obtained with \eq{metric2} and the
co-ordinate transformation \eqs{Atranf}{Btranf}.

The Fourier transforms needed for de-aliasing are carried out using the
Fast Fourier Transform algorithm. While this works for any value of
$N$, it is most efficient when $N$ is an integer power of 2.

We choose our units so that $\kappa^2 =\lambda=1$. Setting $\kappa^2$
defines the energy scale, but on first sight choosing $\lambda=1$ may seem
to result in a loss of generality. However, all terms in the equations of
motion which do not involve $V$ or $V_{,\phi}$ contain two derivatives
with respect to $\eta$ and/or $\zeta$ and the rescaling $\eta\rightarrow
\eta/\sqrt{\lambda}$, $\zeta\rightarrow \zeta/\sqrt{\lambda}$ ensures that
all the terms in the equations of motion are proportional to $\lambda$. 
Consequently, our choice of $\lambda$ is equivalent to a rescaling of time
and length.  All the values of $\zeta$ and $\eta$ given in our results
reflect this scaling.  Finally, when we plot the spectra for variables
such as $\Phi$ and $Q$ we normalize $k$ so that $k=1$ corresponds to the
mode whose wavelength is equal to the {\em initial\/} Hubble radius. 

\subsection{The Perturbative Limit}

Consider the connection between \eq{metric2} and the familiar gauge
invariant theory of cosmological perturbations \cite{MukhanovET1992b}.
In conformal time, the most general metric that describes scalar
perturbations to a FRW metric with flat spatial sections and whose
spatial dependence is restricted to the $\zeta$ direction is
\begin{eqnarray}
ds^2 &=& a(\eta)^2 \{ (1+2\varphi(\eta,\zeta))d\eta^2 -
2b(\eta,\zeta)_{,i} dx^i d\eta \nonumber \\ & & -
\left[(1-2\psi(\eta,\zeta))\delta_{ij} + 2
E(\eta,\zeta)_{,ij}\right]\} dx^idx^j.
\label{pertmetric}
\end{eqnarray}
By comparing this with \eq{metric2} and using the definition of the
gauge invariant perturbation, $\Phi$,
\cite{MukhanovET1992b,ParryET1998a} we obtain
\be
\Phi_{,\zeta\zeta} = \varphi_{,\zeta\zeta} - 
  \frac{1}{a} \frac{d}{d\eta} (E_{,\eta\zeta\zeta}a),
\ee
or
\be
\Phi_{,\zeta\zeta} =  \varphi_{,\zeta\zeta} - 
  \frac{1}{a} \frac{d}{d\eta}\left[ 
 \frac{\alphaT \alpha - \betaT \beta}{a} - H(\alpha^2 -
\beta^2)\right].
\ee
In order to obtain $Q$, we also need the gauge invariant perturbation, 
$\dphigi$. For our metric \cite{MukhanovET1992b},
\be
\dphigi = \delta\phi - \frac{d \phi_0}{d\eta} E_{,\eta} ,
\ee
where $\delta\phi$ is the gauge dependent perturbation measured at
constant $\eta$.  After differentiating twice with respect to $\zeta$,
and again comparing \eq{pertmetric} with \eq{metric2} we find that
\be
\dphigi_{,\zeta\zeta} = \delta\phi_{,\zeta\zeta} - \nonumber \\
\frac{1}{a^2} \frac{d \phi_0}{d\eta} (\alpha\alphaT - \beta\betaT
-\frac{a_{,\eta}}{a} (\alpha^2 - \beta^2)) .
\ee

Our numerical simulations thus give the second derivatives of $\Phi$
and $\dphigi$ with respect to $\zeta$. To compare our results with
perturbation theory, we need the Fourier transforms of $\Phi$ and
$\dphigi$. We find these by transforming $\Phi_{,\zeta\zeta}$ and
$\dphigi_{,\zeta\zeta}$, and multiplying by $-k^2$ which, thanks to
the general properties of the Fourier transform, yields $\Phi_k$ and
$\dphigi_k$. We obtain the background variables ($a(\eta)$,
$\phi_0(\eta)$) by averaging over the spatial grid at each value of
$\eta$.  Once this is done, we have all the raw materials we need to
calculate $Q_k$ from \eq{Qvar}, and make the comparison with
perturbation theory. In models with more than one field, each field
has a gauge invariant perturbation analogous to $\dphigi$.

\begin{figure}[tbp]
\begin{center}
\begin{tabular}{c}
\epsfysize=4cm
\epsfbox{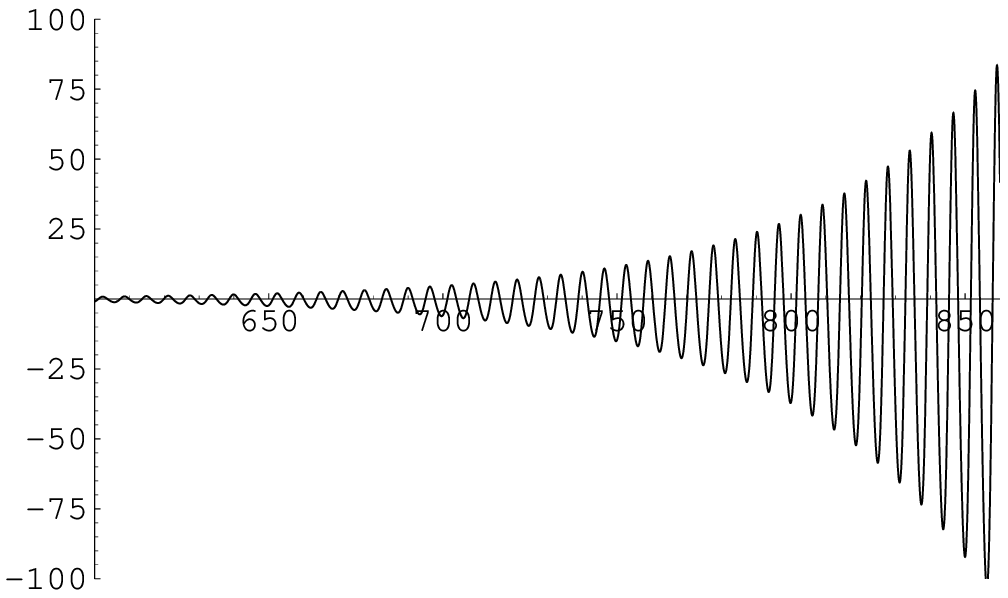} \\
\epsfysize=4cm
\epsfbox{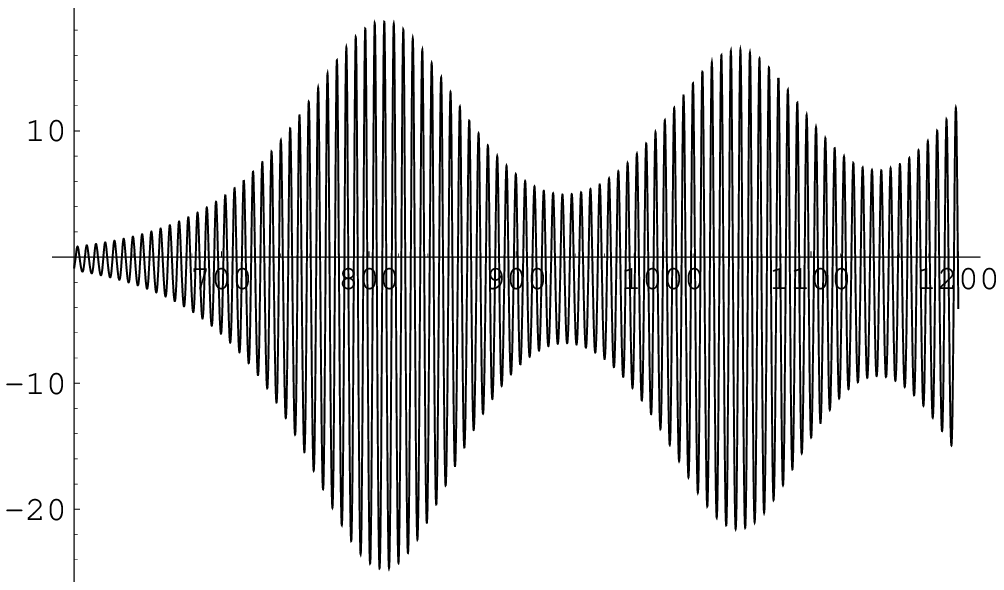} \\
\epsfysize=4cm
\epsfbox{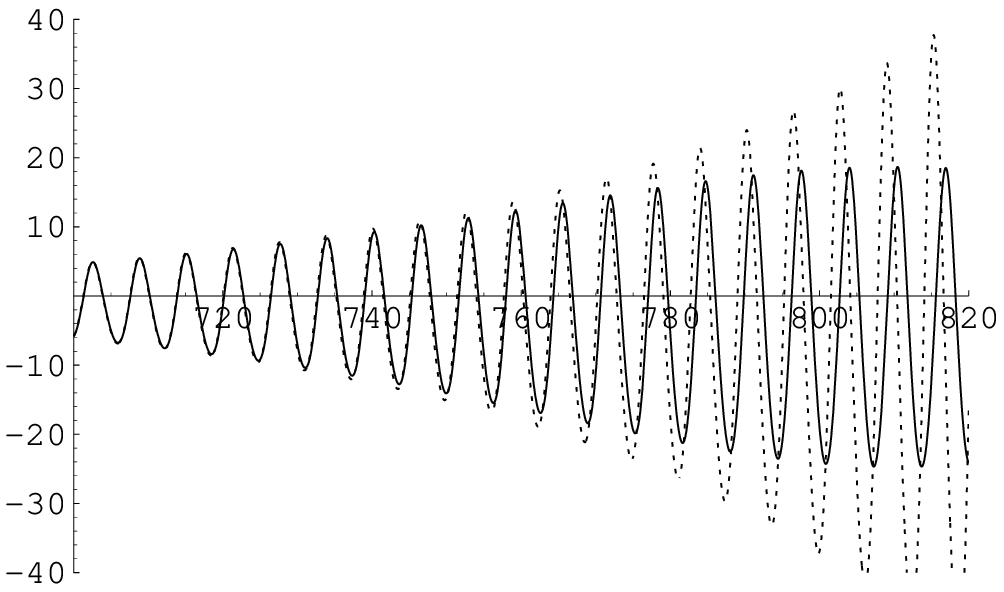} 
\end{tabular}
\end{center}
\caption[]{The evolution of a single resonant mode is plotted as a
function of $\eta$. The top panel shows the evolution of $Q_\phi$
predicted by perturbation theory. The second plot shows the evolution
of $Q_\phi$ derived from the full Einstein field equations. The third
panel superimposes the perturbative and non-perturbative results at
the moment when perturbation theory breaks down. \label{onemode1}}
\end{figure}

\begin{figure}[htbp]
\begin{center}
\begin{tabular}{c}
\epsfysize=4cm
\epsfbox{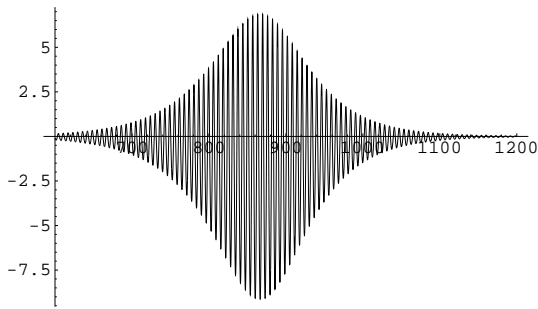} 
\end{tabular}
\end{center}
\caption[]{We plot the evolution of $Q_\phi$ ($\chi/a$) derived from
the nonlinear field equations with a static background.  The specific
mode plotted is the same as that shown in \fig{onemode1}, but the
normalization is different since we dropped the $a''/a$ term before
solving \eq{chieofm}. \label{onemode2} }
\end{figure}

\section{Results}

\subsection{A Single Excited Mode}

We begin our analysis with an initial field field perturbation that
consists of a single mode, chosen to lie within the resonance band.
The parameters for the integration are
\bea
&N=65336=2^{16},\, \delta\eta=\delta\zeta = 0.0722,\, k_{res} = 583&,
\nonumber \\
&\phi_0 = -0.58132,\, \dot{\phi}_0=0 \nonumber
\eea
where $k_{res}$ denotes the mode number of the single excited mode, and
$k=1$ corresponds to the mode whose wavelength is equal in size to the
box. The values for $\phi_0$ and $\dot{\phi}_0$ correspond to the first
turn-around point after inflation. In the plots we normalize $k$ so that
$k=1$ corresponds to a mode with a wavelength equal to the initial Hubble
horizon size, and $k_{res} \sim 1.2$. The initial field values correspond
to the field being at the end of the first oscillation it makes after the
end of inflation.  With this value of $N$, approximately 4000 distinct
Fourier modes can be resolved after de-aliasing. The spacing between the
points, $\delta\zeta$, ensures that the simulation volume initially
contains approximately 500 distinct Hubble volumes. As the comoving
wavelength of the resonant modes is slightly less than the Hubble length
at the end of the inflation we can resolve modes with frequencies up to 7
times higher than the frequencies of modes in the resonance band. 

\Fig{onemode1} contrasts the perturbative evolution of our single excited
mode, found by solving \eq{Qpert}, with the $Q_\phi$ extracted from
the relativistic calculation. Initially, perturbation theory is valid
and $Q$ increases exponentially. While perturbation theory predicts
that exponential growth continues indefinitely, non-perturbative
effects eventually terminate the resonance, as shown in the plots. The
termination of resonance is not attributable to gravitational effects,
and is well known from studies of \eq{chieofm}.\Fig{onemode2} shows
$Q_\phi$ obtained by solving \eq{chieofm}, which assumes that $\Phi=0$
and so $Q$ reduces to $\delta\phi$.  While the resonance terminates,
the subsequent evolution differs from the relativistic result, since
in the absence of gravity the amplitude of the perturbation decreases
after the end of resonance.  Conversely, the $Q_\phi$ obtained from
the full relativistic calculation remains comparatively large after
the resonant growth terminates. Presumably this is due to the
self-gravity of the perturbation, which only enters perturbation
theory at second order.

\begin{figure}[tb]
\begin{center}
\begin{tabular}{c}
\epsfysize=4cm
\epsfbox{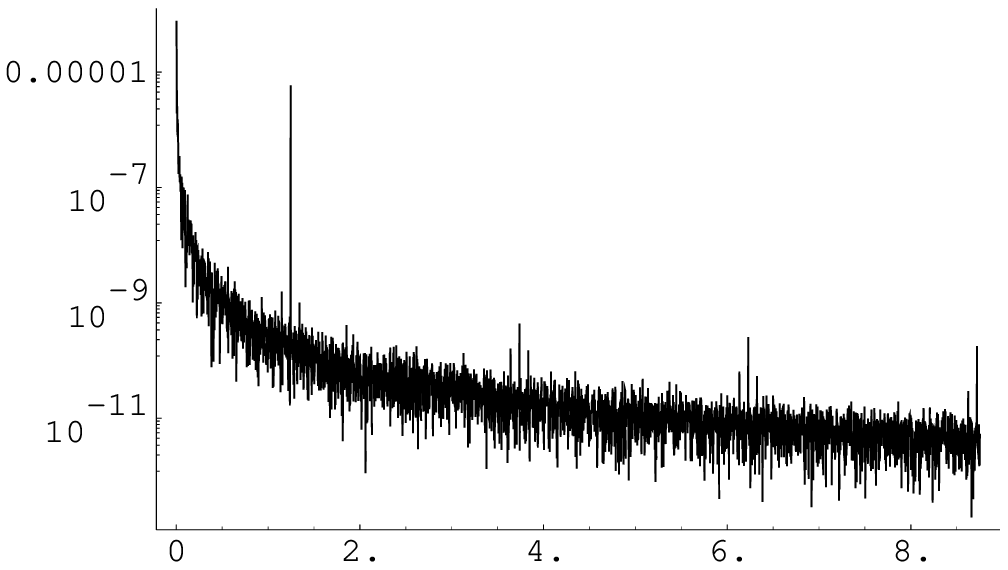} \\
\epsfysize=4cm
\epsfbox{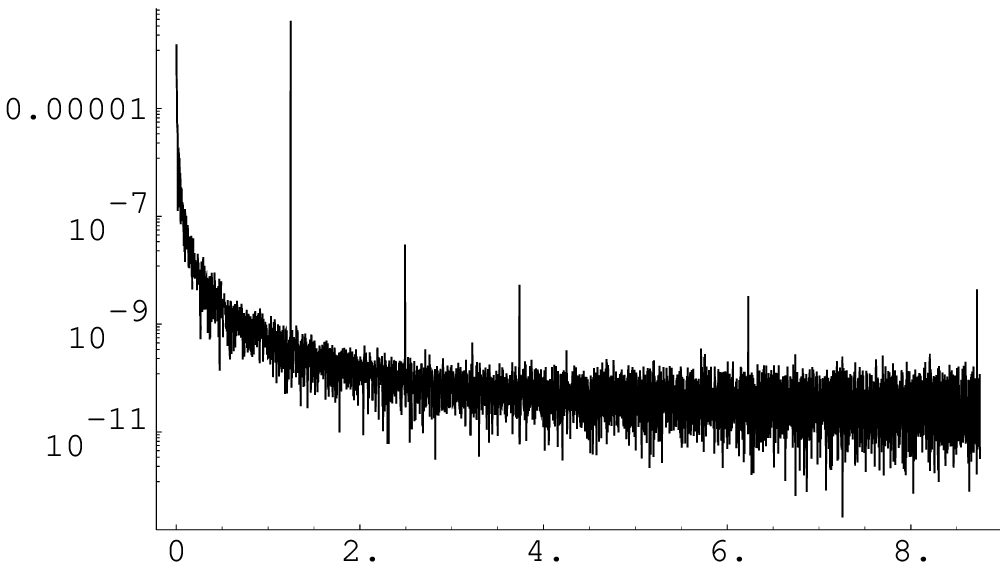} \\
\epsfysize=4cm
\epsfbox{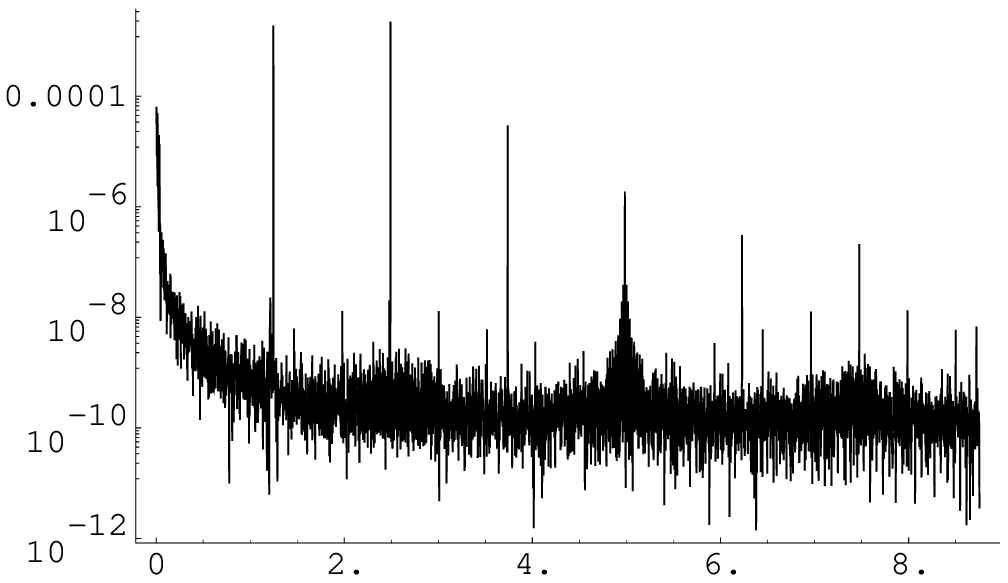} 
\end{tabular}
\end{center}
\caption[]{The power spectrum, $|\Phi_k|$ of the gauge-invariant
metric perturbation is plotted against $k$. A single resonant mode is
initially excited, but since we specify the initial perturbation in a
specific gauge, all modes of $\Phi$ have some initial power.  The
plots show $|\Phi_k|$ at $\eta = 0.22, 414.8$ and $902.8$.  The
$x$-axis is scaled to that a mode with a wavelength equal to the
initial Hubble radius has $k=1$. \label{onemode3}}
\end{figure}

We do not expect the solutions for $\chi/a$ and $\phi$ to overlap
exactly since we have dropped the $a''/a$ term from \eq{chieofm}.
This term quickly becomes irrelevant, but its absence during the
initial stages of the evolution means that a perfect match is not
possible, even with the same initial conditions.  Moreover, this
calculation includes the ``compensating'' field $\psi$, even though it
is not strictly necessary with a single mode and this also changes the
initial normalization of $\phi$, relative to $\chi/a$.

In addition to terminating the resonance, nonlinear effects induce
mode-mode couplings which transfer power to modes outside the
resonance band.  This can be seen in \fig{onemode3}, where we plot the
power spectrum of the gauge invariant perturbation, $\Phi$.  As we
specify the initial perturbation in a specific gauge, the
corresponding gauge invariant quantity actually possesses some power
in all modes. As time progresses, the nonlinear couplings amplify the
``harmonics'' of the resonant mode.  This effect is also apparent when
we solve \eq{chieofm}, but the quantitative results differ from the
general relativistic calculation due to the observed decay of the 
perturbations in the absence of gravity.

\begin{figure}[tbp]
\begin{center}
\begin{tabular}{c}
\epsfysize=4cm
\epsfbox{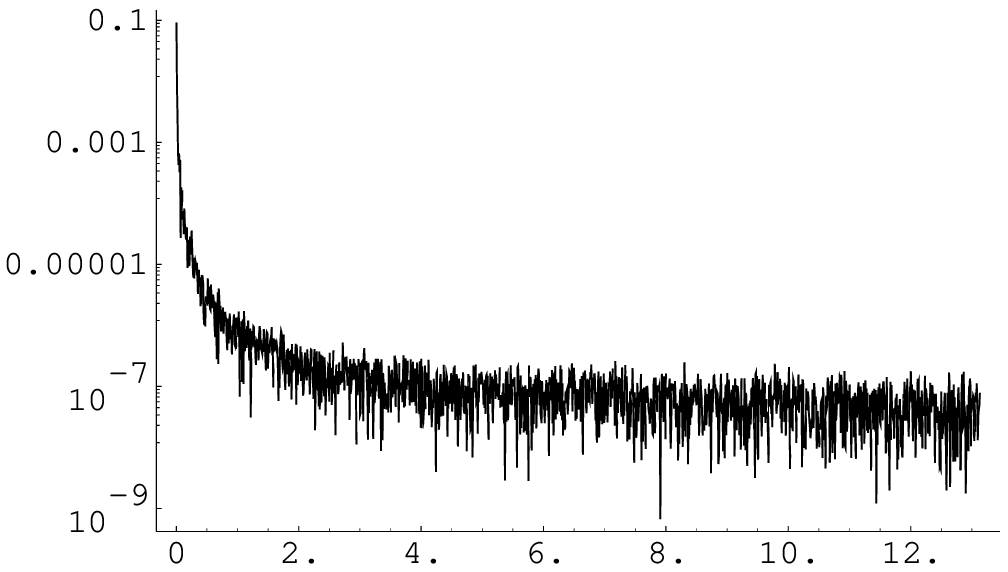} \\
\epsfysize=4cm
\epsfbox{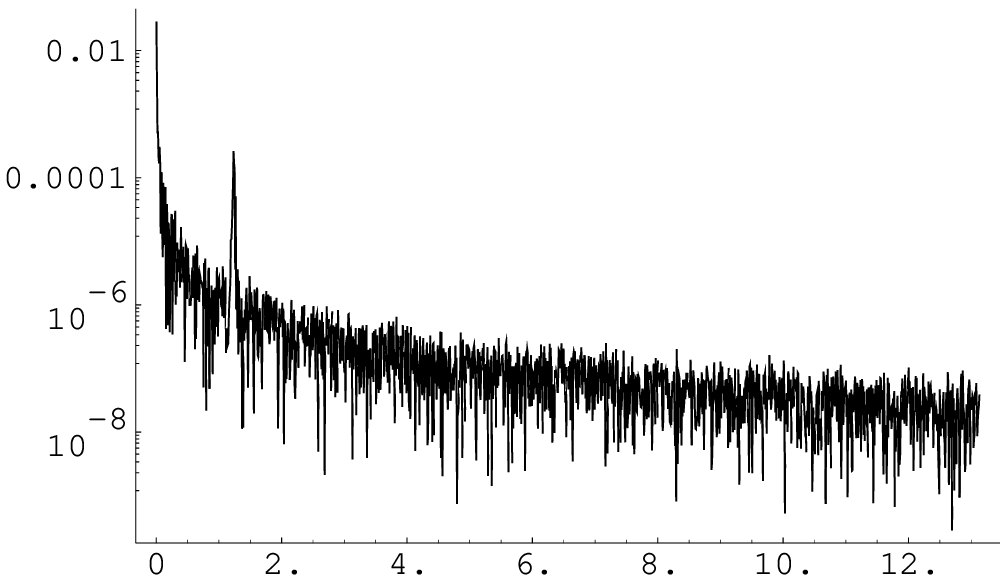} \\
\epsfysize=4cm
\epsfbox{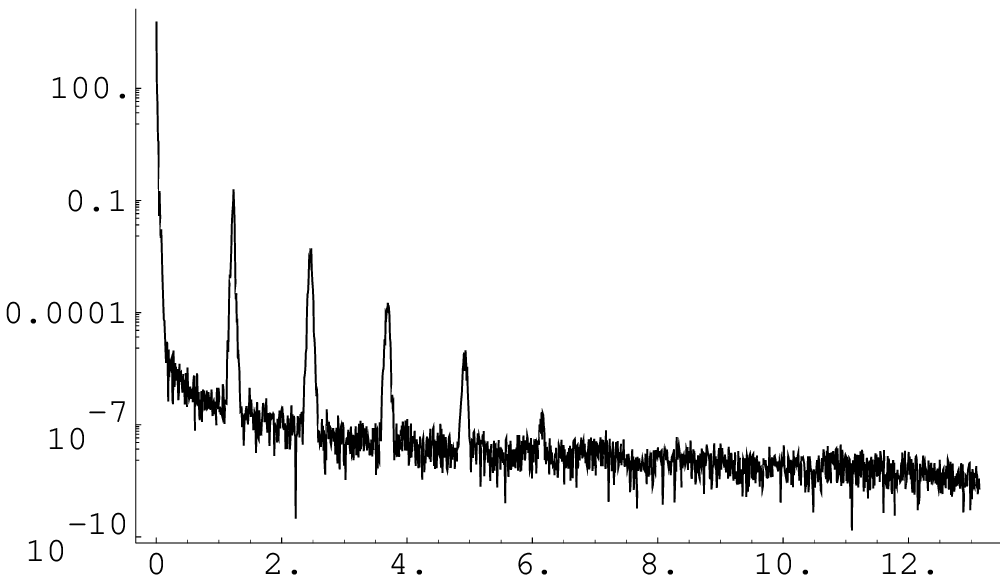} \\
\epsfysize=4cm
\epsfbox{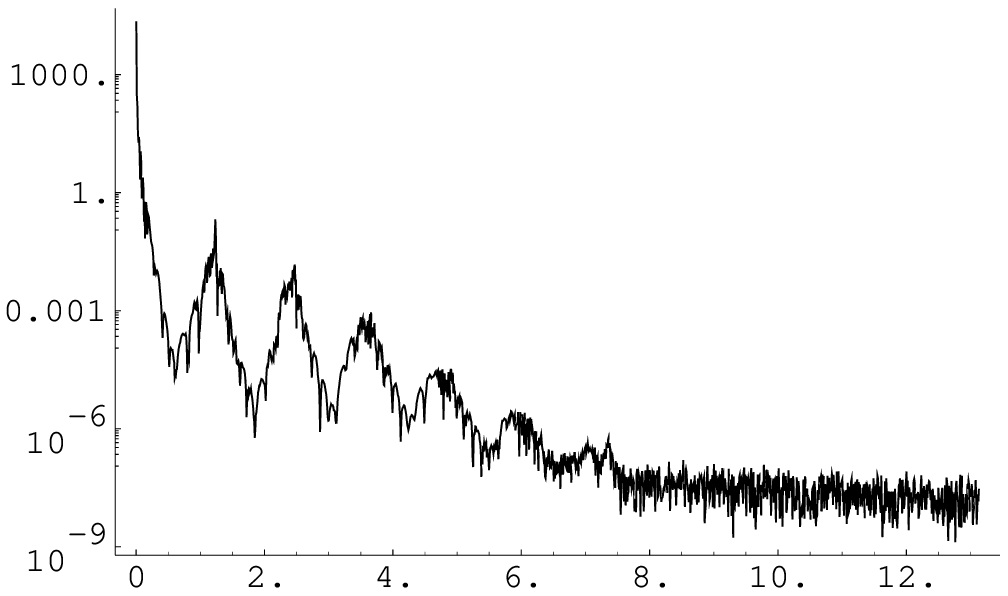}\\
\epsfysize=4cm
\epsfbox{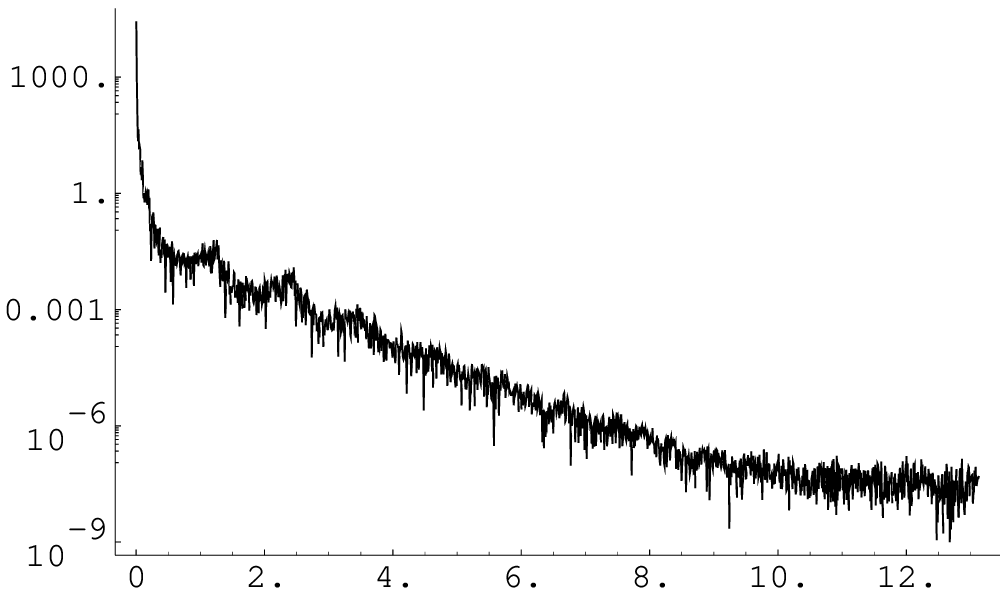}  
\end{tabular}
\end{center}
\caption[]{The power spectrum, $|\Phi_k|$, of the gauge-invariant
metric perturbation is plotted against $k$, with all modes initially
excited.  The spectra are plotted at $\eta=$0.0183, 274.5, 439.2,
750.3 and 915, respectively.  \label{manymode1}}
\end{figure}

\begin{figure}[tbp]
\begin{center}
\begin{tabular}{c}
\epsfysize=4cm
\epsfbox{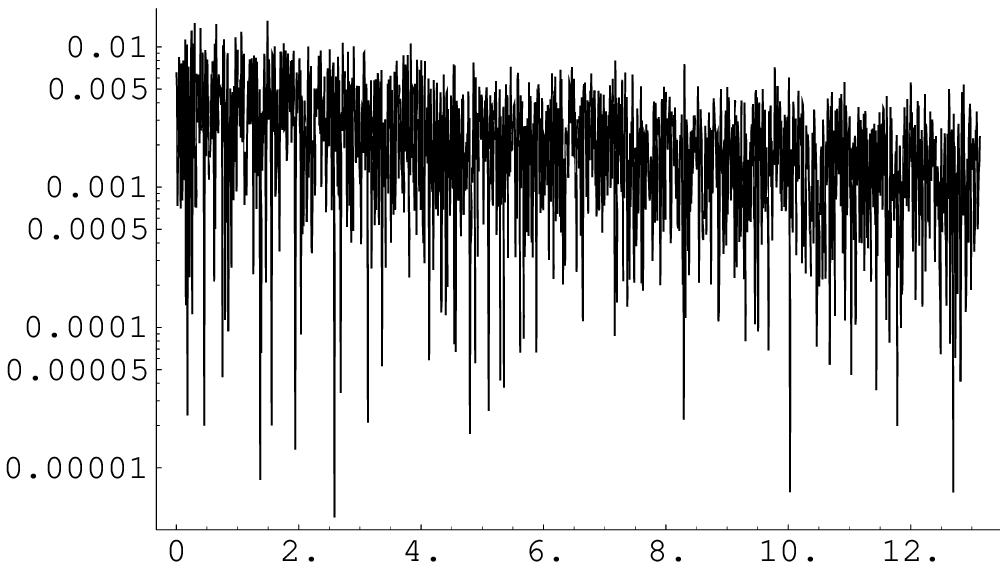} \\
\epsfysize=4cm
\epsfbox{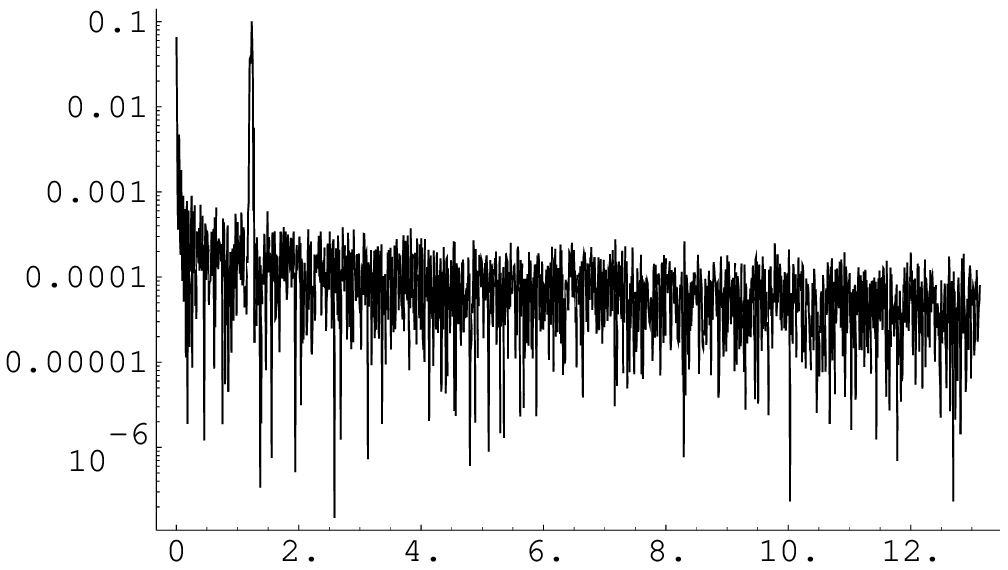} \\
\epsfysize=4cm
\epsfbox{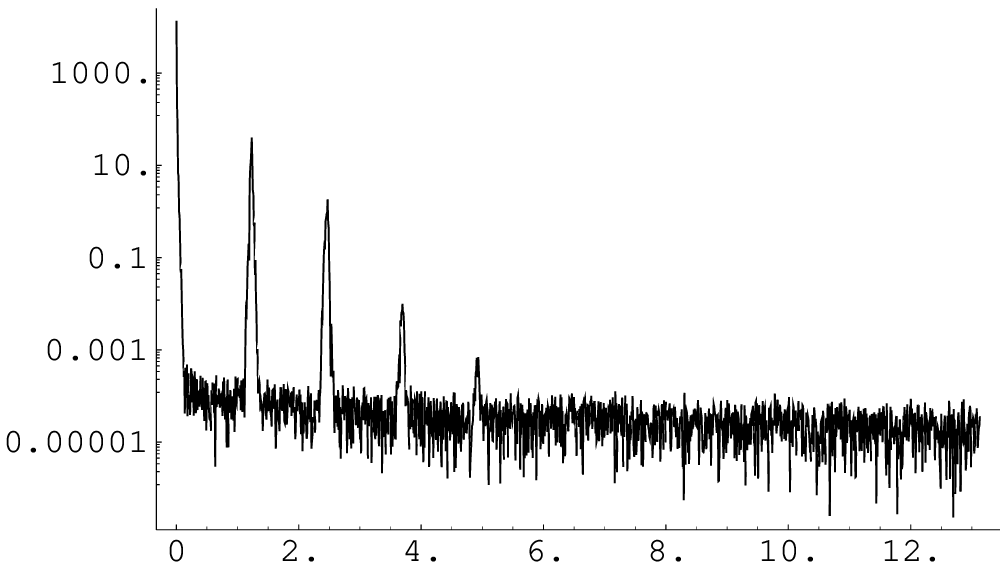} \\
\epsfysize=4cm
\epsfbox{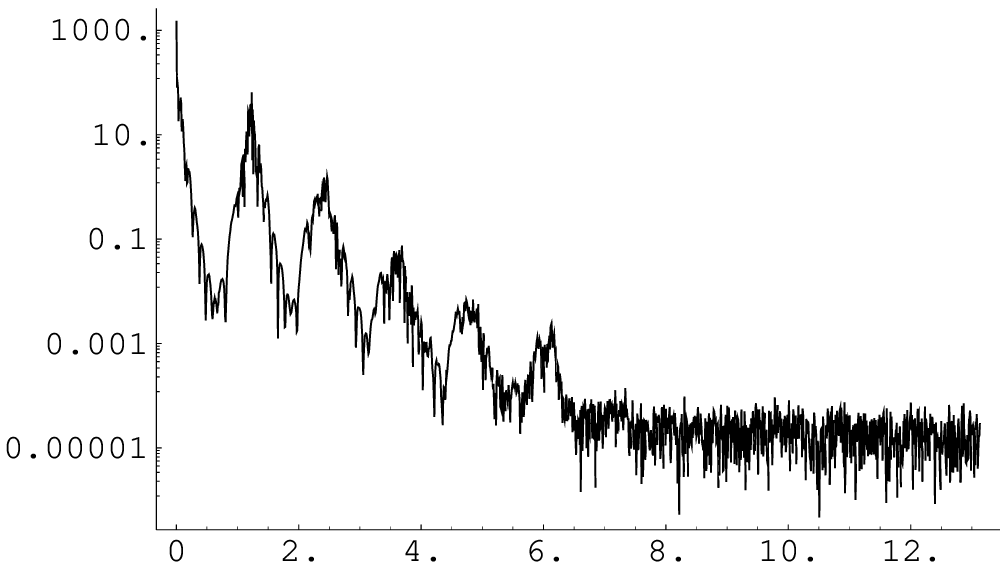}\\
\epsfysize=4cm
\epsfbox{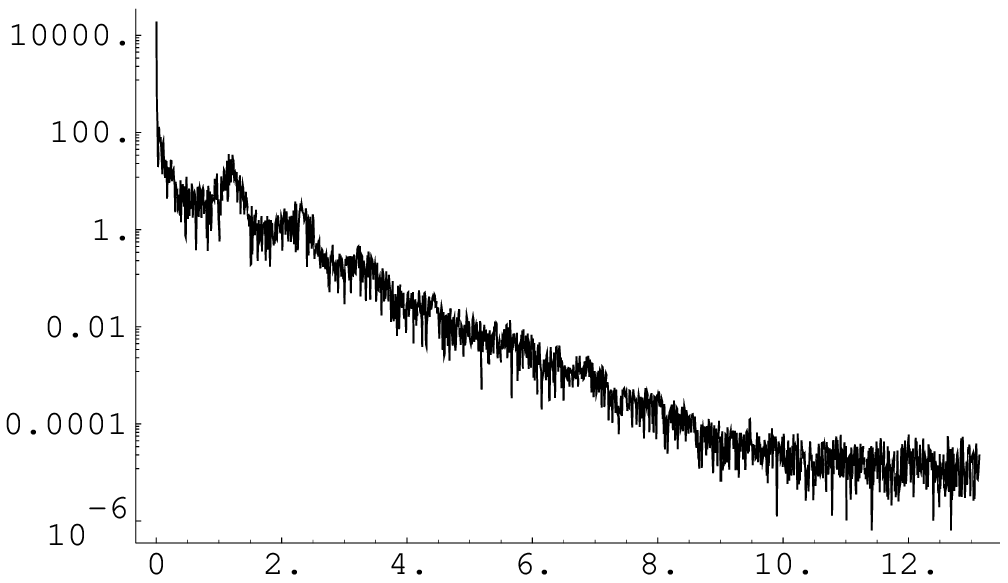}  
\end{tabular}
\end{center}
\caption[]{The power spectrum of $|Q_k|$, is plotted against $k$,
with all modes initially excited.  The spectra are plotted at
the same times as in \fig{manymode1}.
\label{manymode1a}}
\end{figure}

\begin{figure}[tbp]
\begin{center}
\begin{tabular}{c}
\epsfysize=4cm
\epsfbox{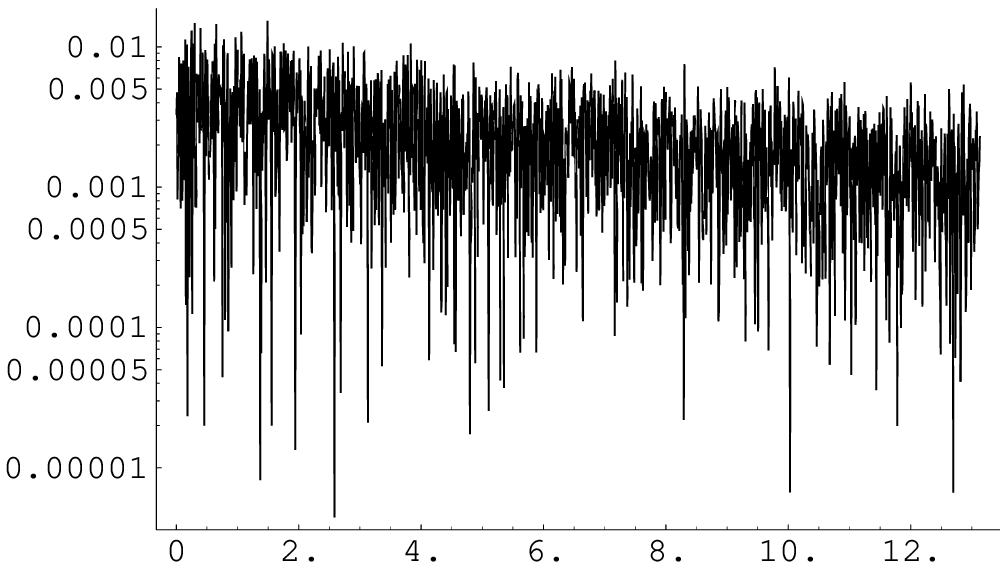} \\
\epsfysize=4cm
\epsfbox{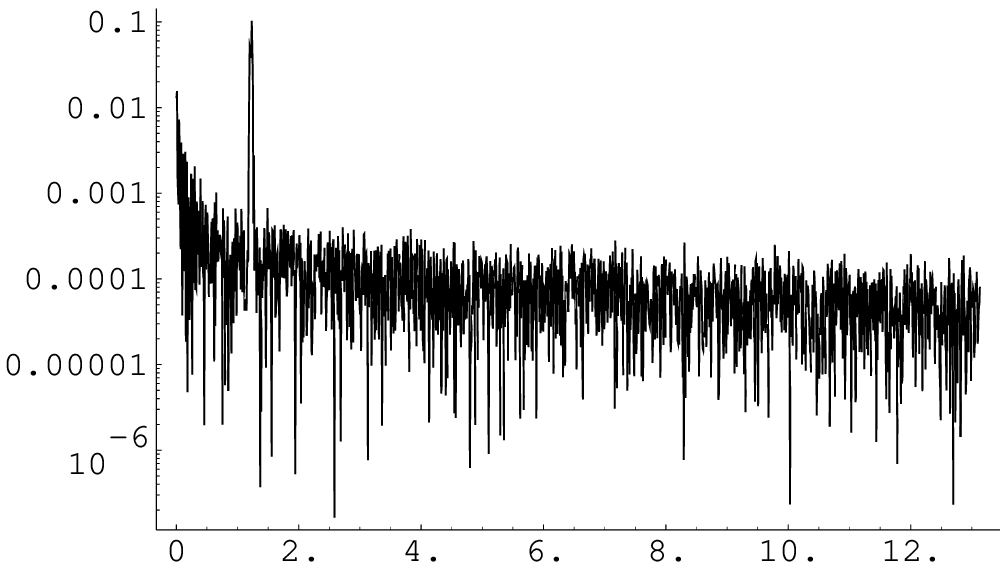} \\
\epsfysize=4cm
\epsfbox{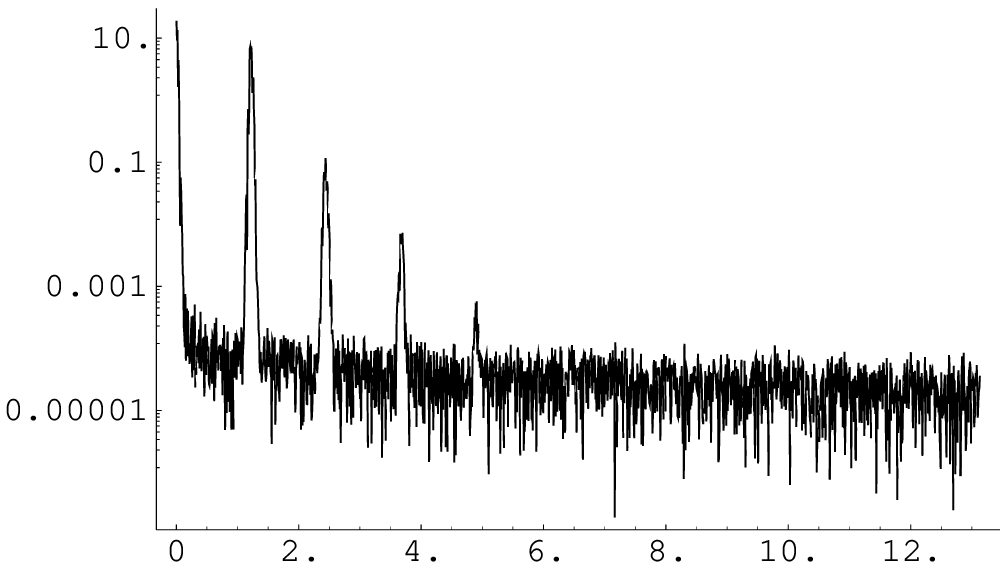} \\
\epsfysize=4cm
\epsfbox{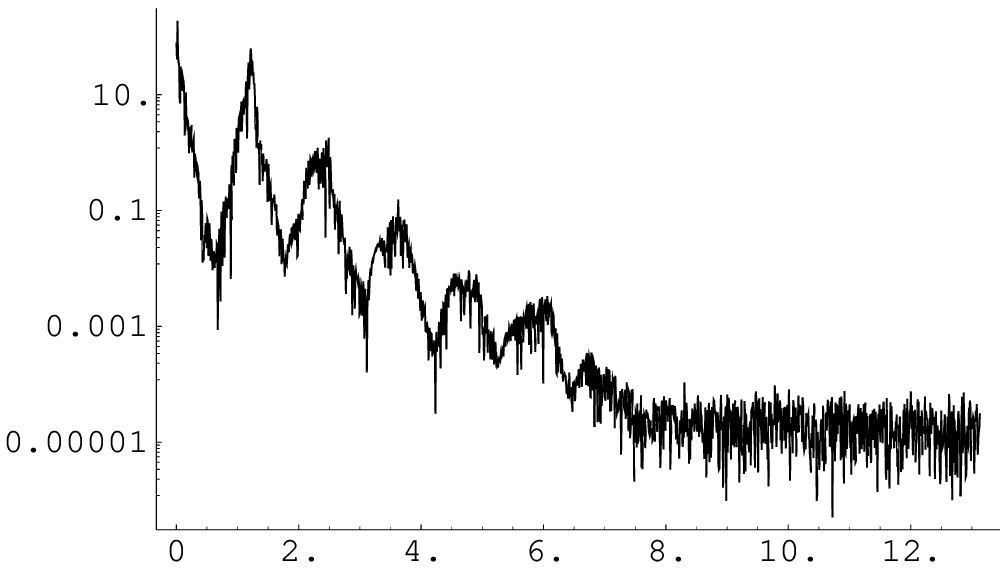}\\
\epsfysize=4cm
\epsfbox{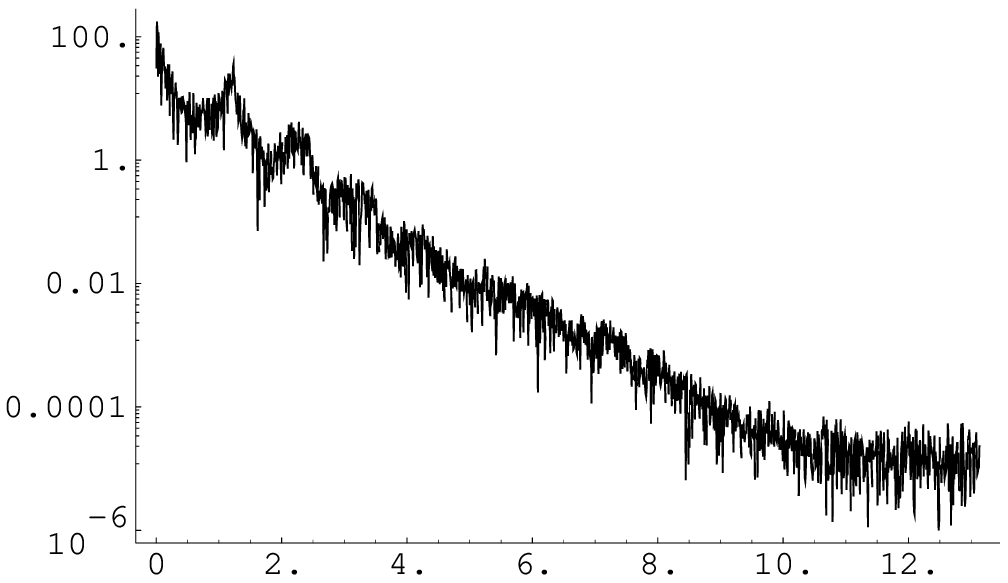}  
\end{tabular}
\end{center}
\caption[]{The power spectrum of $|\chi_k/a|$, derived from a solution of
\eq{chieofm}, is plotted against $k$, with all modes
initially excited.  The spectra are plotted at the same times as in
\fig{manymode1} and \fig{manymode1a}.
\label{manymode1b}}
\end{figure}

\subsection{General Initial Conditions}

Having explored the nonlinear evolution found when a single mode of
the perturbation was initially excited, we turn to the more realistic
case in which the initial amplitude of all modes is non-zero. The
initial amplitudes are set using \eqs{qic1}{qic2}, while the
parameters for the simulation were
\bea
&N=262144=2^{18},\, \delta\eta=\delta\zeta = 0.0183.&
\nonumber \\
&\phi_0 = -0.58132,\, \dot{\phi}_0=0 \nonumber
\eea
The actual volume is the same as in the previous simulation, since the
higher value of $N$ is balanced by the lower value of $\delta\zeta$,
but the spatial resolution is improved.

The amplitude of the initial perturbation here is significantly larger
than the value obtained by extrapolating the COBE normalization to the
scales probed by our simulation.  We made this choice to reduce the
time required for nonlinear effects to become important. However,
apart from prolonging the epoch during which the perturbative
approximation is valid, repeating the calculation with an initial
inhomogeneity that is 100 times smaller does not appear to make any
significant difference to our results.

The most important difference between the multi-mode case and the
previous computation is in the behavior of the longest wavelength
modes. Recall that in the linear approximation (and in the absence of
a transient, decaying solution), and with $k\ll aH$,
\be
 Q \propto \frac{\dot{\phi}}{H}, \qquad 
\Phi \propto 1 -\frac{H}{a} \int_{t_0}^t a(t') dt', \label{smallk}
\ee
so that Q is oscillatory.  With a single excited mode nonlinear
couplings amplified higher harmonics of the resonant mode, but the the
long wavelength (low $k$) modes did not grow, and obey the
relationships in \eq{smallk}.  Conversely, with all the modes
initially excited the longest modes undergo considerable growth,
implying that the linearity assumption of \eq{smallk} is violated, and
that mode-mode effects make a large contribution to the evolution of
the longest modes.  In \fig{manymode1} we plot $|\Phi_k|$ for the
multi-mode case at five different times during the simulation.
Initially, only resonant modes grow significantly.  However, once
nonlinear effects begin to transfer power to the harmonics of the
modes in the resonance band, we see that modes near $k=0$ also start
to grow.  The resonant growth terminates, but the ``bands'' continue
to broaden, and eventually merge to produce a continuous spectrum.
Finally, at the end of resonance the $\chi_k/a$ modes do not decay, as
they did when only a single mode was excited.

We understand the evolution as follows. The initial growth of modes in
the resonance band is explained by the perturbative analysis. The
subsequent enhancement of the harmonics and the modes near $k=0$ is
due to mode-mode couplings. Second order terms transfer power from
modes labeled by $k_1$ and $k_2$ to the modes with $k_1+k_2$ and $|k_1
- k_2|$.  With just one excited mode, $k_1=k_2$, so we can produce the
mode-doubling seeing in the previous simulation. However, for a single
mode $k_1-k_2=0$, and modes with low $k$ are not excited.  Finally,
the broadening of the ``bands'' is (at minimum) a third-order effect.
For instance, given three modes in one of the narrow bands,
combinations like $k=k_1+k_2 -k_3$, transfer power to modes adjacent
to the band. This accounts for the observation that the higher
harmonics are excited well before the bands themselves start to
broaden.

We also note that there is an alternative hypothesis which might
explain the enhancement of the long wavelength modes. A conventional
perturbative calculation predicts that these modes do not undergo
resonance, but this calculation ignores the back reaction of the
resonant perturbations on the oscillating homogeneous mode. This will
induce a subtle change in the forcing term in the perturbation
equations, which is provided by the oscillating field, and this
correction could conceivably induce new resonances at long
wavelengths. This is consistent with our results, since a single
excited mode will have a much smaller back reaction than that produced
by the full spectrum of excited modes, accounting for the absence of
this effect in the single mode simulation. Moreover, the resonance
structure of $\lambda\phi^4$ inflation is known to be very sensitive
to small changes in the driving function. If this is correct, the
harmonics of the resonant modes would still be excited through
mode-mode couplings, but the longest modes would be excited by a
co-operative effect which would involve all the resonant modes. We
stress that we have not yet attempted to verify this hypothesis. Note
too that the possibility of backreaction by perturbations on the
homogeneous mode is discussed (albeit in different models) in
\cite{BassettET1999b,BassettET1999c}.

The most important question is the extent to which these effects are
attributable to nonlinear {\em gravitational\/} couplings, as opposed
to the nonlinear couplings provided by the $\lambda \phi^4$ potential.
Looking at the solutions of $\chi/a$ found via \eq{chieofm}, we see a
similar, but not identical, resonance structure to that predicted by
relativistic calculation.  The discrepancy is most obvious at long
wavelengths. In \fig{manymode1a} we plot $Q_k$ and in \fig{manymode1b}
we show $\chi_k/a$ at the same times as the plots in \fig{manymode1}.

\begin{figure}[tbp]
\begin{center}
\begin{tabular}{c}
\epsfysize=4cm
\epsfbox{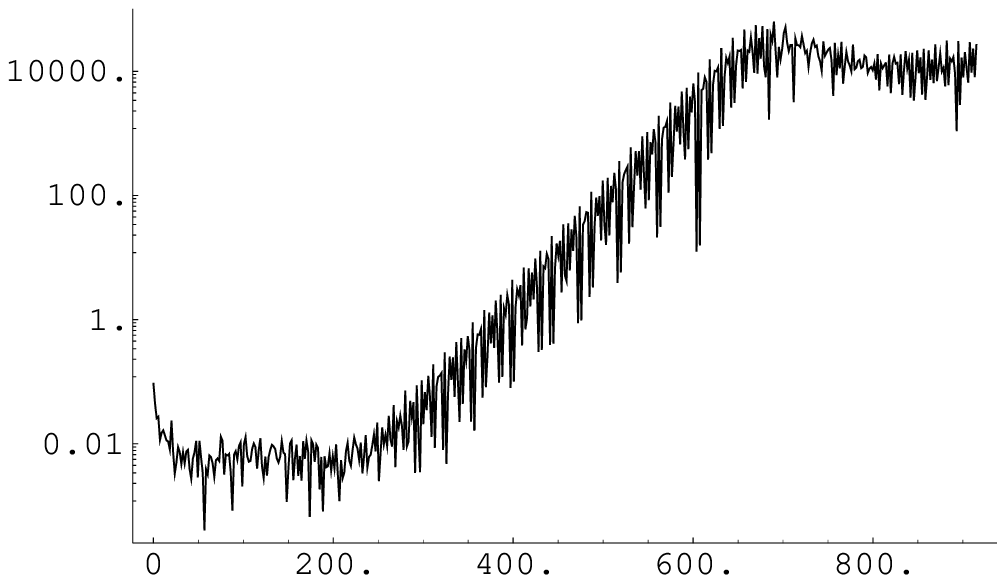} \\
\epsfysize=4cm
\epsfbox{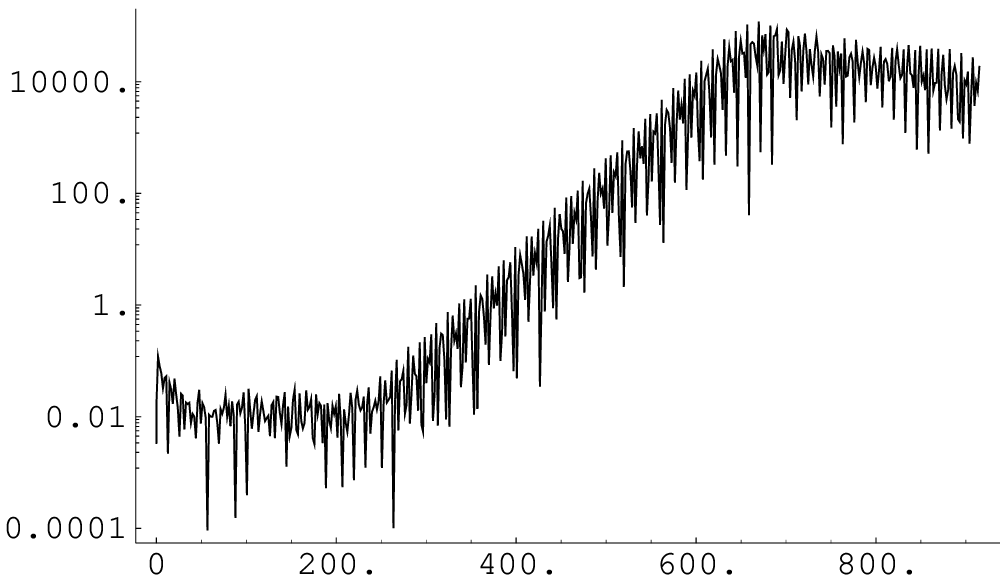} \\
\epsfysize=4cm
\epsfbox{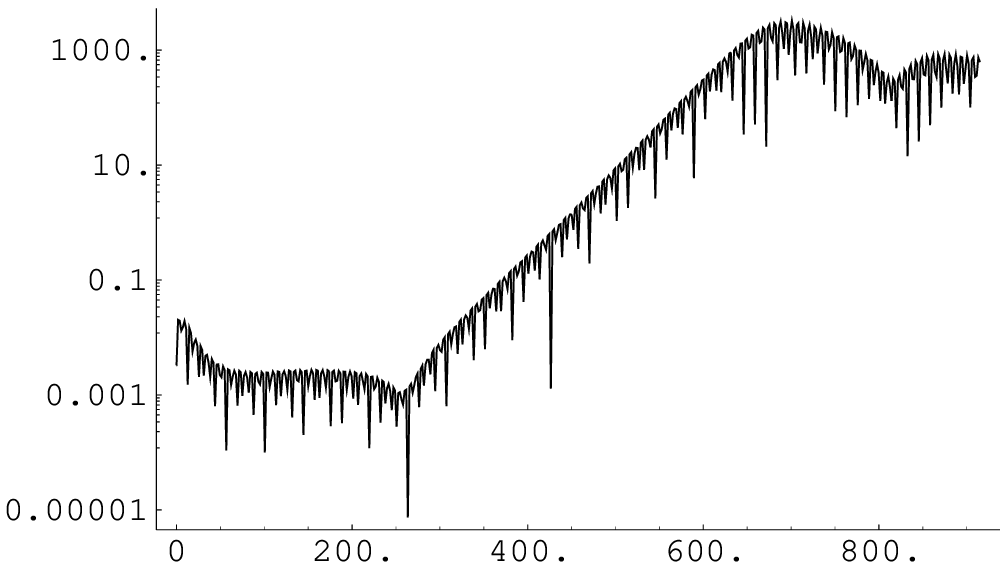} \\
\epsfysize=4cm
\epsfbox{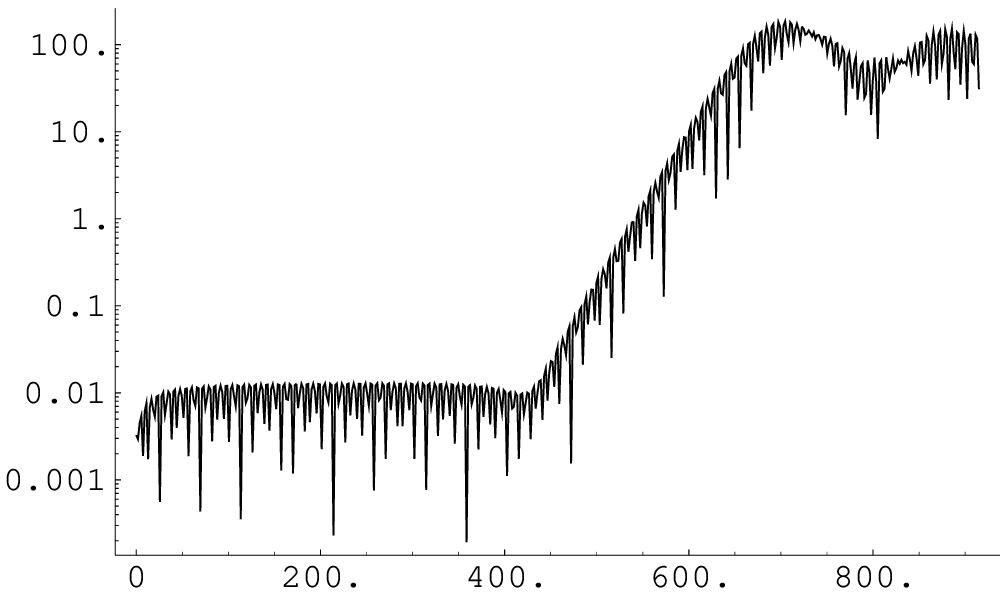} \\
\epsfysize=4cm
\epsfbox{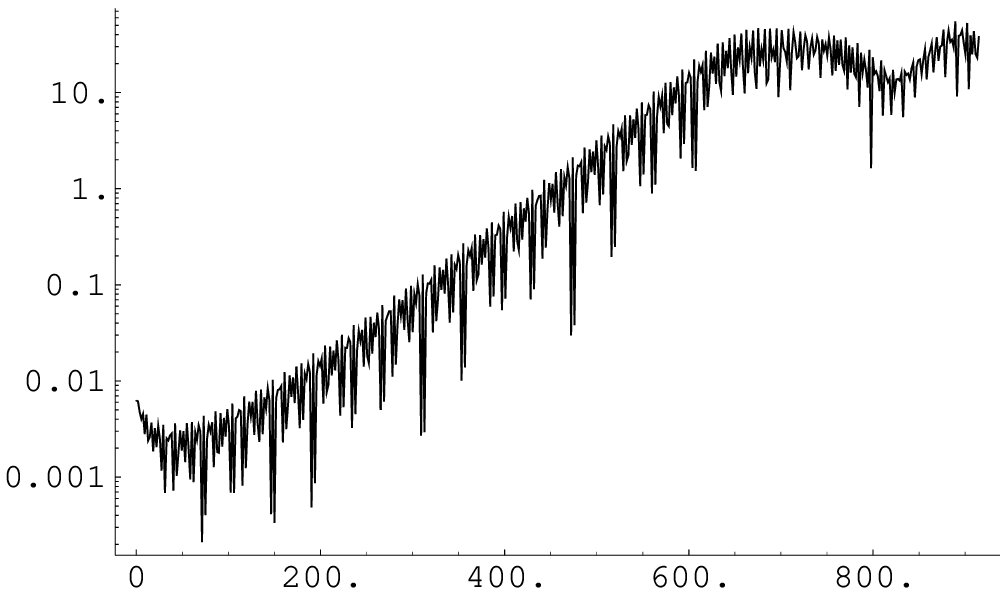} 
\end{tabular}
\end{center}
\caption[]{From top to bottom, these plots show $|\Phi_1|$, $|Q_1|$,
$|\delta \phi_1|$, $|\delta \chi_1/a|$ and $|Q_{583}|$ as functions of
$\eta$. \label{manymode2}}
\end{figure}

\begin{figure}[tb]
\begin{center}
\begin{tabular}{c}
\epsfysize=4cm
\epsfbox{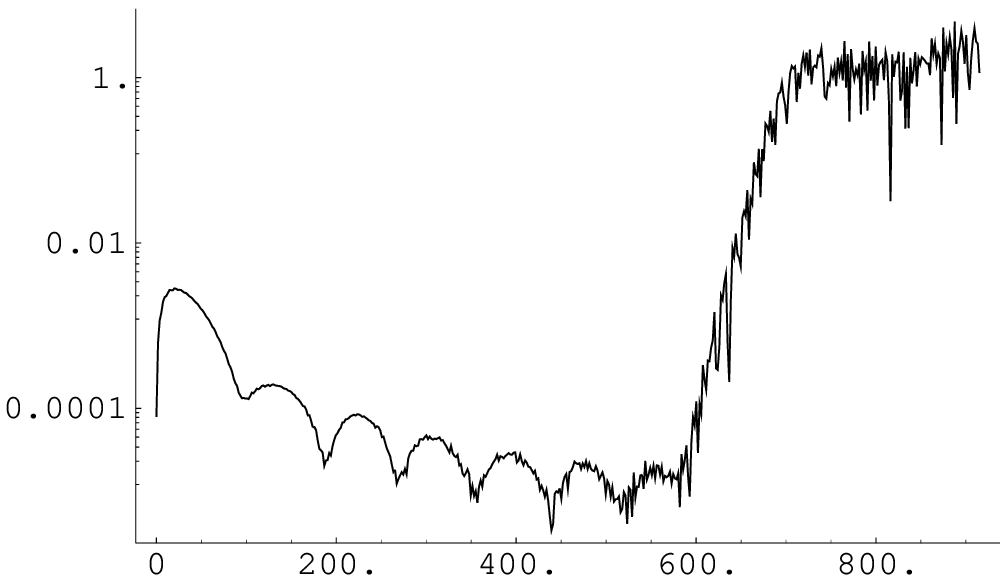} \\
\epsfysize=4cm
\epsfbox{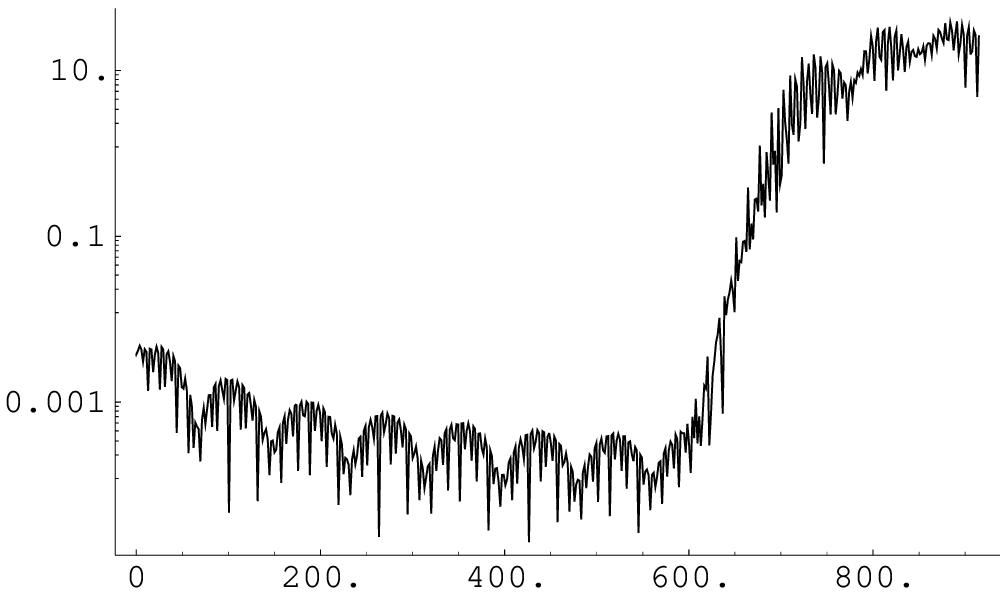} \\
\epsfysize=4cm
\epsfbox{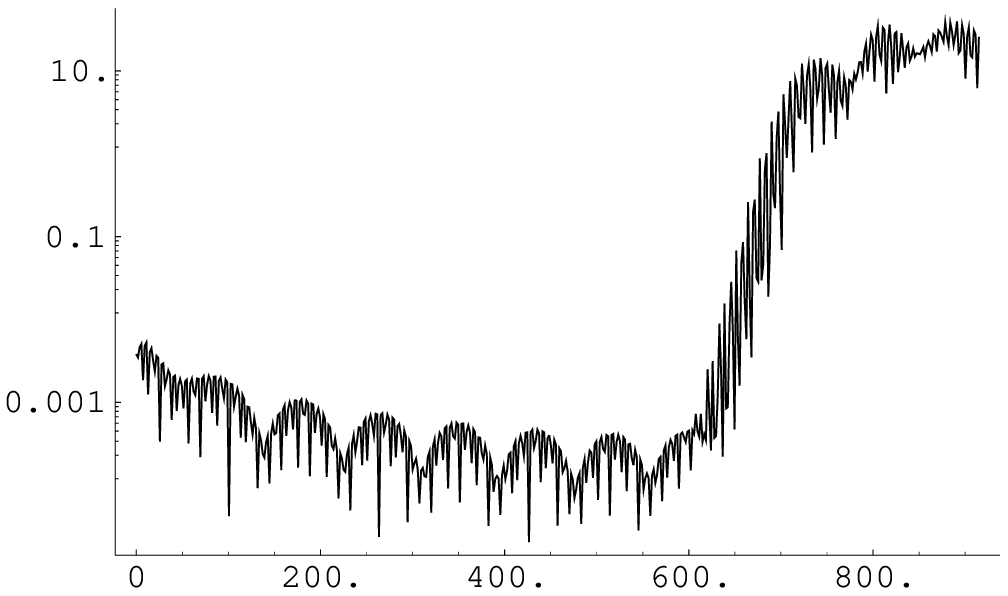} \\
\epsfysize=4cm
\epsfbox{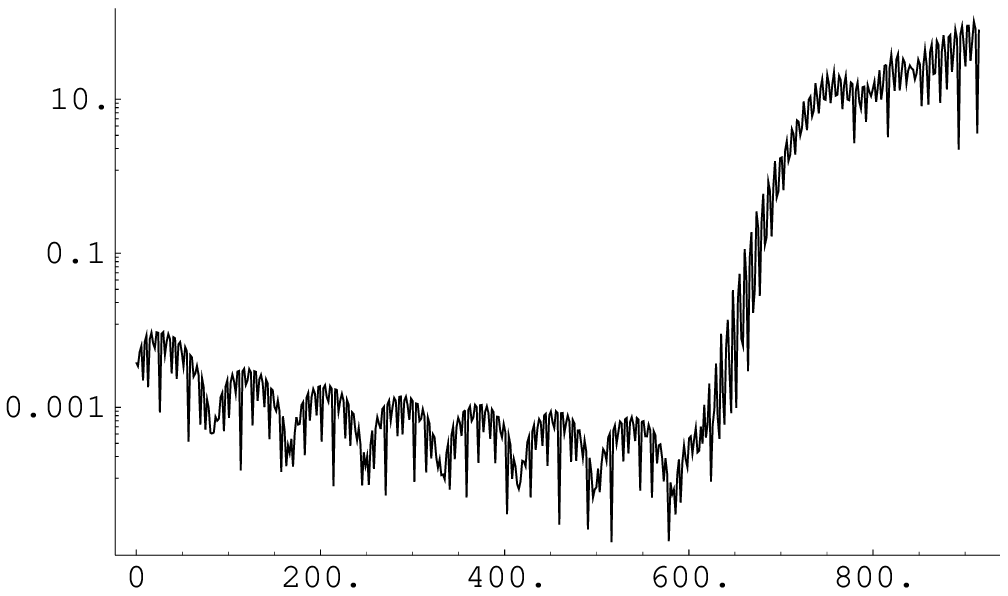} 
\end{tabular}
\end{center}
\caption[]{From top to bottom, these plots show $|\Phi_{50}|$, $|Q_{50}|$,
$|\delta \phi_{50}|$, and $|\delta \chi_{50}/a|$ as function of
$\eta$. \label{manymode3}}
\end{figure}

It is more appropriate to compare $\chi/a$ to $Q$ (or $\dphigi$)
than to $\Phi$, since the latter vanishes if the metric is
unperturbed.  At short wavelengths, the $Q_k$ and $\chi_k/a$ are very
similar, but at the longest wavelengths $Q_k$ and $\dphigi$
undergo much more growth than $\chi_k/a$.  Moreover, while the longest
modes of $\chi_k/a$ {\em do\/} grow, albeit less dramatically than the
corresponding $Q_k$, the growth begins significantly later. This is
illustrated by \fig{manymode2}, which plots the evolution of the
longest modes as a function of $\eta$. Conversely, \fig{manymode3}
shows the same plot for $k=50$, and in this case the difference
between the evolution of $\chi/a$ and the relativistic quantities is
much less pronounced. The final plot in \fig{manymode2} shows the
growth of $Q$ for a mode inside the resonance band (this is actually
the mode that was excited in the single mode case). Looking at this
plot, we see that the near exponential growth of the long-wavelength
modes ceases when the resonance is terminated.

We complete our analysis of \figs{manymode2}{manymode3} by numerically
estimating the slopes of the the exponential segments of the plots.
These slopes contain an arbitrary multiplicative factor, fixed by the
normalization of $a$, which enters via the definition of the conformal
time, $\eta$.  Numerically, and with our normalization of $\eta$, we
find that the exponential segments of the first four plots in
\fig{manymode2} have a slope of approximately $0.036$. Conversely, the
resonant mode shown in the bottom panel has a slope that is half of
this value ($0.017$), supporting our contention that the growth in the
longest modes is due to nonlinear couplings between two adjacent
resonant modes. Finally, the exponential segment of the plots in
\fig{manymode3} has a slope of around $0.08$, confirming that this
growth is driven by a strongly nonlinear multi-mode coupling.

\begin{figure}[tb]
\begin{center}
\begin{tabular}{c}
\epsfysize=4cm
\epsfbox{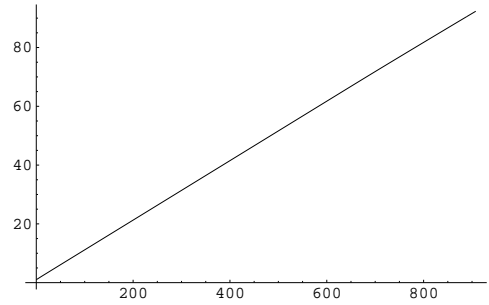} 
\end{tabular}
\end{center}
\caption[]{This plot shows the scale factor $a(\eta)$, obtained by
averaging the metric functions obtained from our simulation to remove
the perturbation.   \label{aeta}}
\end{figure}

\begin{figure}[tbp]
\begin{center}
\begin{tabular}{c}
\epsfysize=4cm
\epsfbox{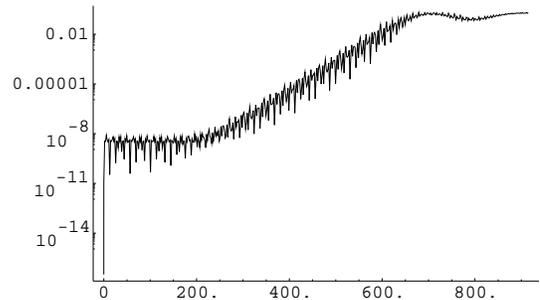} 
\end{tabular}
\end{center}
\caption[]{The growth of ${\cal{C}}^2/{\cal{R}}^2$ during the course of
the simulation is plotted as a function of $\eta$.   \label{weyl2}}
\end{figure}

\begin{figure}[tbp]
\begin{center}
\begin{tabular}{c}
\epsfysize=4cm
\epsfbox{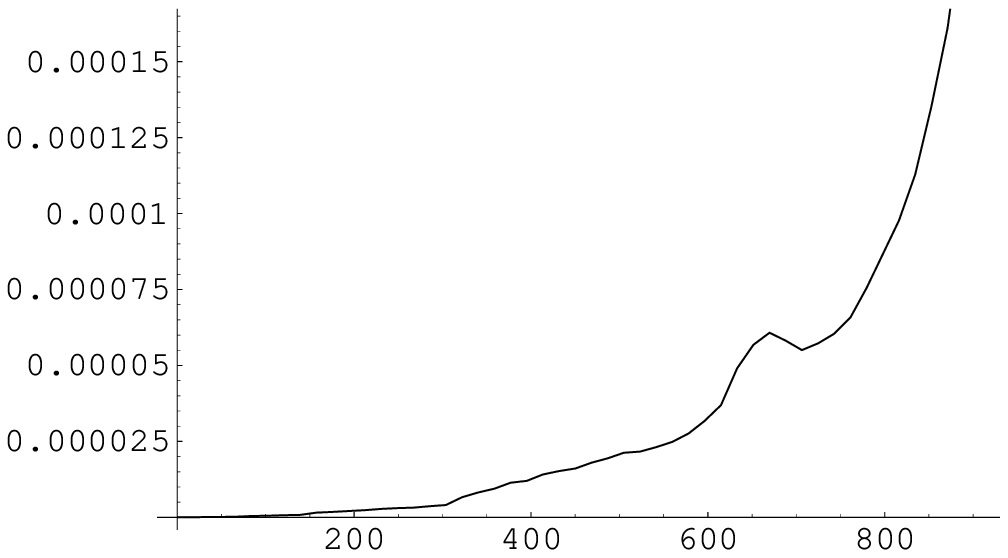} 
\end{tabular}
\end{center}
\caption[]{The evolution of $\gamma$, \eq{gamma}, which quantifies the
accuracy with which we satisfy the constraint, \eq{constr22a}, is
plotted for our numerical solution.  For an exact solution $\gamma$
would be identically zero.  \label{constraint} }
\end{figure}

Since we are working with the full Einstein field equations we can
examine non-perturbative measures of curvature. For instance, $C^2 =
C_{\mu\nu\gamma\lambda}C^{\mu\nu\gamma\lambda}$, where $C$ is the
usual Weyl tensor, is a covariant scalar density that is sensitive to
the local spacetime curvature. Consequently, we can define
\bea
{\cal{C}}^2(\eta) &=& \int{d\zeta \sqrt{-g} C^2(\zeta,\eta)} \\
{\cal{R}}^2(\eta) &=& \int{d\zeta \sqrt{-g} R^2(\zeta,\eta)}
\eea
where $R$ is the usual curvature scalar. These integrals are not
covariant, since they depend on a specific co-ordinate
choice. However, ${\cal{C}}^2/{\cal{R}}^2$ is dimensionless, and gives
a qualitative measure of the inhomogeneity on our hypersurfaces, which
we plot in \fig{weyl2}, because $C^2$ vanishes everywhere in an
FRW spacetime. Thanks to our 
choice of initial conditions,
the metric functions are constants at $\eta=0$, so $C^2$ vanishes on
the initial hypersurface. Hence, the magnitude of $C^2/R^2$ is a
measure of how far our spacetime departs from a perfectly FRW universe. The
initial entropy perturbation quickly generates a metric
perturbation. This remains constant until the nonlinear growth of the
longest modes sets in at around $\eta=200$, whereupon
${\cal{C}}^2/{\cal{R}}^2$ grows exponentially until the underlying
resonance is halted. However, after resonance ends
${\cal{C}}^2/{\cal{R}}^2$ remains roughly constant.

If the enhanced density perturbations generated during resonance led
to localized gravitational collapse and the formation of primordial
black holes (or ``black walls'' in the case we consider here), this
would be obvious from ${\cal{C}}^2/{\cal{R}}^2$ as the curvature would
diverge in a collapsing region. Since this does not happen in our
simulation we find no evidence that preheating after $\lambda \phi^4$
inflation leads to primordial black hole formation. However, this
result should be interpreted with care. After inflation, the equation
of state resembles a relativistic gas, where $p=\rho/3$, and as a
consequence gravitational collapse is much less efficient than in dust
filled universe with $p=0$.\footnote{Recall the absence of structure
formation in a hot dark matter dominated universe, as compared to a
cold dark matter universe with the same primordial perturbation
spectrum.}  Consequently, even if preheating does sometimes lead to
primordial black holes, it would not be surprising that they were
absent from the $\lambda\phi^4$ case. In future work we will study
more general potentials, and will be able to consider this topic in
greater detail.

Apart from the length scale provided by the resonance band, the other
important scale in the post-inflationary universe is the Hubble scale,
$H^{-1}$.  However, this length is not constant during the course of the
simulation. In \fig{aeta} we plot the scale factor corresponding to
the unperturbed FRW limit of our simulation, and we see that the
universe grows 90 times larger during the simulation. During the same
interval the Hubble radius grows linearly from an initial value of 10
to a final value of 900 (in our units). Our simulation region has a
comoving width of 4797 ($2^{18}\delta\zeta$), so the $k$-th mode has a
comoving wavelength $4797/k$. Modes with $k\lesssim 500$ are initially
outside the horizon. At the time when the $k=1$ ($\eta \sim 200$) mode
begins to grow, only modes with $k\lesssim25$ are outside the Hubble
horizon. By the end of the simulation, only four or five modes have
wavelengths greater than $H^{-1}$.

Modes with $k \gtrsim 50$ show few differences between the evolution
of $Q_k$ obtained from the relativistic calculation, and the analogous
result for $\chi_k/a$. Consequently, on these scales (which can safely
be termed ``sub-Hubble'') we see no evidence that nonlinear
gravitational effects influence the overall progress of preheating
after $\lambda \phi^4$ inflation.

For the longest modes there are clear differences between the results
found in a fixed FRW background and the full relativistic calculation.
Both the relativistic calculation and the nonlinear field equations in
an unperturbed background predict that the longest modes of the field
perturbation grow significantly. However, the relativistic results
predict that this growth begins earlier and is an order of magnitude
larger than that found from the field evolution in a rigid
background. Note that in both cases we are seeing the growth of modes
that lie outside the instantaneous Hubble horizon.

At present, we do not identify the source of this discrepancy, but
several possibilities exist. The most interesting explanation would
obviously be back reaction, or some other nonlinear effect
\cite{BassettET1998a,BassettET1999a,MukhanovET1997b,AbramoET1997a}.
However, we stress that we have no evidence that the specific results
of our simulations are predicted by these authors.

There are a number of less glamorous possible explanations for the
enhanced inhomogeneity, which we briefly review. Firstly, it could be
a side effect of using periodic boundary conditions, which induce an
artificial long-range correlation into our simulation.  Similarly, we
assume that the inhomogeneity is limited to one spatial dimension, and
the nonlinear effects we observe may be an artifact of this
restriction.  In many nonlinear systems the dimensionality has a
crucial influence on the behavior, so we would still need to verify
that any cosmologically significant changes to long wavelength modes
persisted in the fully inhomogeneous case.  On the other hand, our
calculation of $\chi/a$ is also restricted to a single inhomogeneous
dimension, so the dimensional dependence of $Q$ and $\chi/a$ would
have to be different if this explanation is correct.  Likewise, the
three dimensional simulations of Khlebnikov and Tkachev
\cite{KhlebnikovET1996a}, show a band structure that is qualitatively
similar to that obtained here for $\chi/a$, which suggest that at
shorter wavelengths our results are reliable. However, since we
consider with a $1+1$ dimensional system we are able to resolve a
large range of length-scales, whereas the simulation region in
Khlebnikov and Tkachev's work is not significantly larger than the
initial Hubble volume. Consequently, we cannot compare our calculation
to theirs at super-Hubble scales.

We also need to be sure that the results we have seen do not depend on
our introduction of the $\psi$ field to satisfy the initial
constraint. Ideally in future work we will solve the constraints
directly on the initial surface and dispense with $\psi$ entirely.
However, we have varied the initial fraction of $\psi$ and checked
that changing it made no significant difference to our results.

A different possibility is that our ``reference'' hypersurfaces,
defined as the 3-planes with constant $\eta$, are degenerate in some
way. If this is true the degeneracy must be shared by $t(\eta,\zeta)$
and $z(\eta,\zeta)$, which we compute from \eqs{tft}{tfz}, as $z$ and
$t$ remain smooth functions of $\zeta$ and $\eta$ during the
simulation. Moreover, we have been careful to work with variables
which are invariant (to first order) under gauge transformations.

Numerically, we checked our solutions by ensuring that they did not
change significantly when we altered the parameters that govern the
numerical integration scheme. In particular, the results appear to be
independent of $N$, the number of points on our grid, the distance
between them, $\delta\zeta$, and the timestep $\delta\eta$. The
accuracy with which we fit the constraints provides an independent
test of our results, and the constraints are indeed well satisfied,
especially during the era when nonlinear effects first become
apparent.

The constraints, \eqs{constr12a}{constr22a}, allow us to form an
independent measure of the accuracy of our code. To measure how well
our solution obeys a given constraint we evaluate
\be
\gamma^2 = \frac{\int{d\zeta(l-r)^2}}
  {\sqrt{\int{d\zeta l^2}}\sqrt{\int{d\zeta r^2}}}.  \label{gamma}
\ee
We have two different $\gamma$, corresponding to the two constraints,
while $l$ and $r$ are the left and right hand sides of the equation
\eq{constr12a} or \eq{constr22a} (depending on which constraint we are
testing), and $\gamma$ is the average fractional difference between
the left and right sides of the constraint equations. In
\fig{constraint} we plot the $\gamma$ derived from \eq{constr22a}, and
we see that it is always small.  Moreover, $\gamma$ decreases with
$\delta\eta$ and $\delta\zeta$. In general, the value of $\gamma$
derived from \eq{constr12a} is smaller than that obtained from
\eq{constr22a}, and we do not show it here.

Our choice of $N$ and $\delta\zeta$ must satisfy two opposing
criteria. Ideally, we would like to be able to resolve modes which
were much larger than the Hubble radius for the entire duration of
resonance, but to prevent the build up of numerical error we must
resolve all the harmonics of the resonant modes that are excited by
the nonlinear couplings. This sets an upper limit on $\delta\zeta$ and
$\delta\eta$.  Including super-Hubble modes sets a lower limit on the
total size of the box. The combination of these two effects is to
ensure that $N$ must be quite large. Consequently, while it is
feasible to include modes which are a few times larger than the Hubble
radius at the end of resonance, looking at modes 100 times larger (or
more) would require a significant amount of computer time.  A better
approach may be to look for similar effects in models where the
resonance is stronger, which means that the overall growth of the
universe during the resonant era will be much smaller than in the
$\lambda\phi^4$ case.

\section{Conclusions}

We have analyzed parametric resonance after $\lambda \phi^4$ inflation
for two distinct sets of initial conditions. In the first we gave only
one Fourier mode of the perturbation a non-zero initial amplitude, and
in the second all modes were initially excited.  The overall picture
of the resonant era derived from our calculations, which incorporates
nonlinear gravitational effects, is similar to that obtained from
analyses of the nonlinear field dynamics which assume that the
background spacetime is rigid. However, in both simulations we saw new
phenomena which appear to be due to nonlinear gravitational effects.
With a single mode excited, the perturbation calculated in a rigid
spacetime background decayed after the resonant growth terminated, but
its amplitude after resonance remained large when nonlinear
gravitational effects were included. When all modes were initially
excited power was transferred to long wavelength modes by second order
effects in both the rigid background and nonlinear gravitational
calculations, but in the latter case the overall growth is larger and
begins sooner.  However, we do not see any evidence for the formation
of primordial black holes and the long wavelength modes cease growing
once the exponential increase of the modes inside the resonance band
comes to a halt.

Given the recent interest in nonlinear gravitational effects during
preheating and in gravitational back reaction, the enhanced
inhomogeneity we see at long wavelengths when all modes are initially
excited is a tantalizing result. However, more work will be needed to
confirm that these effects have a physical origin, and are not an
artifact of our numerical scheme or underlying assumptions. Moreover,
if the effect is real, its precise origin and physical consequences
remain to be determined.  In addition, the symmetry condition we have
imposed in our simulation can be shown to preclude the existence of
tensor perturbations to the metric, so we cannot reach any conclusions
about the resonant production of gravitational waves after inflation.

Finally, the numerical codes and analytic techniques developed during
this investigation are readily applicable to other resonant models,
including those based on two (or more) interacting scalar fields.
Since the structure of the resonance after $\lambda \phi^4$ inflation
is very simple, we can form a clear picture of the evolution of the
inhomogeneous parts of the field and the spacetime metric during the
resonant era.  Consequently, in addition to its intrinsic interest,
$\lambda\phi^4$ is an excellent model to begin with as it has allowed
us to perfect the techniques we will use in the analysis of other,
more complicated, systems. This work is currently in progress, and its
results will shed further light on the potential nonlinear
gravitational effects we have uncovered here.

\section*{Acknowledgments} We thank Robert Brandenberger, Fabio
Finelli and Andrew Sornborger for useful discussions.  Computational
work in support of this research was performed at the Theoretical
Physics Computing Facility at Brown University.  RE and MP were
supported by DOE contract DE-FG0291ER40688, Task A.

\end{document}